\documentclass[review,3p]{elsarticle}
\usepackage[utf8]{inputenc}
\usepackage{amsmath,amssymb,amsfonts}
\usepackage{graphicx}
\usepackage{url}
\usepackage{textcomp}
\usepackage{xspace}
\usepackage{booktabs}
\usepackage{subcaption}
\usepackage{fnpct}
\usepackage{pgfplots}
\pgfplotsset{compat=1.14}

\usepackage{algorithm}
\usepackage{algpseudocode}

\newdefinition{definition}{Definition}

\usepackage{tabularx}

\journal{Journal of Systems and Software}

\graphicspath{{./figures/}}

\newcommand{\method}{Tetraband\xspace}
\newcommand{\etal}{\textit{et al.}\xspace}

\newcommand{\B}{\ensuremath{\mathcal{B}}\xspace}
\newcommand{\stc}{\ensuremath{\mathcal{S}}\xspace}
\newcommand{\futc}{\ensuremath{\mathcal{F}}\xspace}

\begin{document}

\begin{frontmatter}

\title{Adaptive Metamorphic Testing with Contextual Bandits}

\author[1]{Helge Spieker\corref{cor1}}
\ead{helge@simula.no}

\author[1]{Arnaud Gotlieb}
\ead{arnaud@simula.no}

\address[1]{Simula Research Laboratory, P.O. Box 134, 1325 Lysaker, Norway}

\cortext[cor1]{Corresponding author}

\begin{abstract}
    Metamorphic Testing is a software testing paradigm which aims at using necessary properties of a system under test, called metamorphic relations, to either check its expected outputs, or to generate new test cases. 
    Metamorphic Testing has been successful to test programs for which a full oracle is not available or to test programs for which there are uncertainties on expected outputs such as learning systems.
    In this article, we propose \textit{Adaptive Metamorphic Testing} as a generalization of a simple yet powerful reinforcement learning technique, namely contextual bandits, to select one of the multiple metamorphic relations available for a program. 
    By using contextual bandits, Adaptive Metamorphic Testing learns which metamorphic relations are likely to transform a source test case, such that it has higher chance to discover faults.
    We present experimental results over two major case studies in machine learning, namely image classification and object detection, and identify weaknesses and robustness boundaries. Adaptive Metamorphic Testing efficiently identifies weaknesses of the tested systems in context of the source test case.
\end{abstract}

\begin{keyword}
Software Testing \sep Metamorphic Testing \sep Contextual Bandits \sep Machine Learning
\end{keyword}

\end{frontmatter}

\section{Introduction}
Metamorphic Testing (MT) is a software testing paradigm that aims at using necessary properties of a software under test to either check its expected outputs or to generate new test cases \cite{Chen1998,Chen2018}.  
More precisely, MT tackles the so-called \textit{oracle problem} which occurs whenever predicting the expected outputs of a system is just too difficult or even impossible. 
Typical examples include machine learning models used for classification tasks, for which only stochastic behaviors can be specified \cite{Barr2015}. 
Indeed, these models are often initially trained with existing datasets and then exploited to classify new data samples. 
However, the expected class of any new data sample is unknown and thus, these samples cannot be used for testing the trained models. 
Fortunately, transformations over the data samples which do not change their (unknown) class, are usually available. 
By applying these transformations, called Metamorphic Relations (MRs) in MT, it becomes possible to effectively test trained machine learning models \cite{Murphy2008,Ding2017,Dwarakanath2018}. 

MT has been very successful to address testing issues in various application domains, e.g., driverless cars \cite{Zhou2019}, search engines \cite{Zhou2016b}, or bioinformatics \cite{Shahri2019} just to name a few (see Section~\ref{sec:relwork} for more references).
However, generally speaking, applying MT in practice requires to address two issues: the \textit{MR identification} and the \textit{MR selection} problems \cite{Chen2018}. 
The former occurs when trying to identify MRs for a specific system, i.e., to formalize input transformation properties which lead to a known transformation of the outputs. 
Finding such relations may be difficult when there is no obvious symmetries in the input data, or obvious system invariant, or else when the functional behavior of the system is unknown. 
The second occurs when several MRs have been identified, but determining which ones are best suited to discover faults in the system under test is hard. 
It is important to select appropriate MRs for testing the system to avoid redundancies in test cases and thus to avoid slack in the test execution process. 
This problem is especially critical when testing is part of a continuous integration process where there is usually a limit on the time allocated for testing in an integration cycle.  

This paper addresses exclusively the latter problem, i.e., MRs selection, by formulating the effective selection of MRs as a reinforcement learning problem, based on \emph{contextual bandits}. 
Our method, called \textit{Adaptive Metamorphic Testing} (AMT), defines a \emph{test transformation bandit} which sequentially selects a MR that is expected to provide the highest payoff, i.e., that is most likely to reveal faults.
Which MRs are likely to reveal faults is learned from successive exploration trials.
The bandit explores the different available MRs and evaluates the fault landscape of the system under test, thereby providing valuable information to the tester.

Learning the selection of MRs can be useful when testing under resource-constraints, for example in cases where the system under test changes are frequently integrated and tested, but also for infrequent testing when the number of MRs is large or their checking is costly.
We also discuss a second application which is to identify robust boundaries of the MR parameters.
Robust boundaries describe related scenarios but focus on a single MR that can be controlled via parameters.
The interest is to find parameters that produce fault-revealing test cases with minimal changes only.

In this paper, we evaluate Adaptive Metamorphic Testing on two case study applications for image analysis, namely, \textit{image classification} and \textit{object detection}.
As implementations of these case studies, we test freely available and pre-trained deep learning systems that can be used as black-box components in other software systems.
For each system, we explore both the general fault-revealing capabilities of metamorphic relations and the discovery of robustness boundaries.

The main contributions of this paper are three-fold.
First, we introduce Adaptive Metamorphic Testing as a general adaptive selection method for metamorphic relations and test transformations.
The method is based on reinforcement learning with contextual bandits and learns to identify those relations which are likely to reveal faults in the system under test. 
This method is useful in a context where a source test case can be modified by several different metamorphic relations and the system is to be repeatedly tested. 
To the best of our knowledge, it is the first time that reinforcement learning is applied to select MRs and embedded into a general methodology for MT.

Second, we provide an implementation of Adaptive Metamorphic Testing in a tool called \method, dedicated to testing machine learning models for image analysis. 
Our tool facilitates metamorphic relations based on image augmentation functions and provides dedicated environments for adaptive metamorphic testing that can be integrated with other implementations.

Third, we explore the benefits of \method on two case studies coming from image analysis, namely image classification and object recognition.
Both of these case studies are relevant subsystems in a wide number of applications, such as autonomous cars, robot navigation, or industrial automation \cite{LeCun2015,Pouyanfar2018}. 
For these applications, high-quality standards are essential and rigorous testing is a requirement.
Our experiments show that \method is highly beneficial to optimize the testing process towards fault-revealing MRs.

The remainder of the paper is structured as follows.
We review the background on metamorphic testing and contextual bandits and related work in the area in Section~\ref{sec:background}. 
Section~\ref{sec:method} introduces Adaptive Metamorphic Testing and its components.
We discuss general application scenarios in Section~\ref{sec:applications} before introducing the experimental setup, consisting of our implementation of AMT, \method, and the two case studies in Section~\ref{sec:experiments}.
The results are presented in Section~\ref{sec:results} and Section~\ref{sec:conclusion} finally concludes the paper.

\section{Background} \label{sec:background}

\subsection{Metamorphic Testing} \label{sec:mettest}

Metamorphic Testing (MT) aims at using necessary properties of a software under test to either check its expected outputs or to generate new test cases \cite{Chen1998,Chen2018}.
Central to MT is the concept of \emph{Metamorphic Relations} (MRs) which are high-level observable properties that must hold over inputs and outputs of the system under test.

In the following, we formalize the definition of a metamorphic relation, the transformation from source test cases to follow-up test cases, and metamorphic testing.
Our definitions follow the formalization by Chen \etal \cite{Chen2018}, except for the transformation from a source to a follow-up test case.
We interpret the transformation function to apply to the whole test case, which includes the test input, and formalize it accordingly in this more general, but compatible, way:

\begin{definition}[Metamorphic Relation (MR)]
Let $f$ be a target function or algorithm.
A metamorphic relation (MR) is a necessary property over a sequence of multiple inputs $\langle x_1, x_2, \dots, x_n \rangle$ ($n \geq 2$) and their corresponding outputs $\langle f(x_1), f(x_2), \dots, f(x_n) \rangle$.
It can be expressed as a relation $\mathcal{R} \subseteq X^n \times Y^n$, with $\subseteq$ being the subset relation, and $X^n$ and $Y^n$ being the Cartesian products of $n$ input and output spaces.
\end{definition}

\begin{definition}[Transformation from Source Test Cases to Follow-up Test Cases]
Consider a MR $\mathcal{R}(x_1, x_2, \dots, x_n, f(x_1), f(x_2), f(x_3))$.
The sequence of inputs and their corresponding outputs defines the set of source test cases.
For each source test case $\stc(x) \in \mathcal{R}$, a follow-up test case $\futc$ is derived by applying a possibly non-deterministic transformation function $T$ to the input of the source test case $x$: $\futc(x) = f(T(x))$.
The transformation function $T$ is constructed such that the follow-up test case $\futc$ fulfills the necessary property of $\mathcal{R}$.
\end{definition}

\begin{definition}[Metamorphic Testing (MT)]
Let $P$ be an implementation of a target algorithm $f$.
For an MR $\mathcal{R}$, suppose that we have $\mathcal{R}(x_1, x_2, \dots, x_n, f(x_1), f(x_2), f(x_3))$.
Metamorphic testing (MT) based on this MR for $P$ involves the following steps:
\begin{enumerate}
    \item Define $\mathcal{R}'$ by replacing $f$ by $P$ in $\mathcal{R}$.
    \item Given a sequence of source test cases $\langle x_1, x_2, \dots, x_k \rangle$, execute them to obtain their respective outputs $\langle P(x_1), P(x_2), \dots, P(x_k) \rangle$. Construct and execute a sequence of follow-up test cases $\langle x_{k+1}, x_{k+2}, \dots, x_{n} \rangle$ according to $\mathcal{R}'$ and obtain their respective outputs $\langle P(x_{k+1}), P(x_{k+2}), \dots, P(x_n) \rangle$.
    \item Examine the results with reference to $\mathcal{R}'$. If $\mathcal{R}'$ is not satisfied, then this MR has revealed that $P$ is faulty.
\end{enumerate}
\end{definition}

For the remainder of this paper, we refer in general to a metamorphic relation $R$ as the tuple of $\langle \mathcal{R}, T \rangle$, the combination of necessary properties over the outputs for a specific MR and the transformation function $T$ to generate follow-up test cases from source test cases.
Therefore, a source test case \stc produces output $P(x)$ for the system $P$ with test input $x$.
Using a metamorphic relation $R$, the follow-up test case \futc with test input $T(x)$, where $T$ is a transformation over the input $x$, can be generated.
Due to the metamorphic relation over $R$ (including $T$), \stc, and \futc, the result $P(T(x))$ can be verified.
Note that MRs are only partial properties, which means that only test cases that violate them indicate the presence of faults in the system under test. 
Showing that the system satisfies a MR on any input or a test suite does not guarantee the absence of faults but increases our confidence in the system correctness.
However, this issue concerns any software testing method, not only MT.  

The transformation function $T$ of a metamorphic relation $R$ does not have to be a deterministic function, but is usually parameterized and can result in several different follow-up test cases from one source test case, depending on a parameter $\phi$ which specifies the exact transformation to be applied.
In many cases and the simplest implementation of metamorphic testing, $\phi$ is chosen from a random distribution and $T$ produces a random follow-up test case.
We use $R_\phi$ to denote a metamorphic relation configured by $\phi$, which makes it deterministic, and $R$ for the general metamorphic relation, which might be non-stochastic if no additional configuration is possible.

\subsection{Contextual Bandits} \label{sec:bandits}

The selection of a test transformation to apply on a source test case is formalized as a multi-armed bandit problem with context information, also known as a contextual bandit \cite{Langford2007,Zhou2016}.
A contextual bandit acts in discrete iterations, where each iteration corresponds to the generation and execution of one follow-up test case.
A bandit \B has $k$ arms, where every arm corresponds to the selection of one possible MR to generate the follow-up test case.
In every iteration $i$, the bandit receives the context vector $c_i$ which in our case describes the source test case.

The bandit then acts according to one or multiple policies $\pi$ which formalize the action selection, i.e. the decision making strategy.
These policies are trained from external feedback about the success of previously made decisions.
A policy is often realized by function approximation techniques, for example, multiple linear regression or neural networks.
Additionally, an exploration strategy is used to try previously unexplored actions instead of following the policy only.
The bandit chooses an arm ($a_i = \pi(c_i)$) and receives a reward, also called payoff, $r_{i}$ which is the external feedback for this decision.
As only one arm can be selected, there is also only feedback for the effect of this single arm $a_i$.
Afterwards, the bandit updates its policy from the observation $(c_i, a_i, r_i)$.
Updating the bandit works by adjusting a set of weights, such that the new set of weights better fits the previously made experiences and minimizes regret for historical decisions.
The actual implementation of the weight update is dependent on the specific contextual bandits algorithm \cite{Agarwal2014} and the learner used to approximate the policy, e.g. whether it is a form of linear regression or a non-linear neural network.

Our goal for the contextual bandit is to maximize the total payoff, i.e., the cumulative undiscounted reward over all iterations $\sum_{i=1}^{T} r_i$.
To achieve this goal and to identify highly rewarded actions, contextual bandit algorithms are designed to also minimize the regret.
The regret of a bandit is the gap between the expected reward over a number of iterations when following one policy and the cumulative reward the agent actually receives over the same number of iterations \cite{Lattimore2019}.
As a smaller regret means to choose actions more closely to the highest possible payoff, minimizing the regret implies the maximization of the total payoff, but by maintaining the concept of regret also badly rewarded actions contribute to the improvement of the policy.

A challenge in the design of contextual bandits is to find a balance between \textit{exploration}, i.e., evaluating the effect of rarely used actions, and \textit{exploitation}, i.e., repeating those actions that showed to be effective before. This process is called the \emph{exploration-exploitation trade-off}.
To this extent, several exploration techniques have been developed as part of bandit algorithms.
We introduce here two techniques that are useful in Adaptive Metamorphic Testing.
\emph{Epsilon-greedy} \cite{Sutton2018} decides to explore a random action with probability $\epsilon$ and with probability $1-\epsilon$ the current learned policy is used to select an action. 
The parameter $\epsilon \in [0,1)$ is chosen by the user.
A more advanced exploration strategy is \emph{online cover} \cite{Agarwal2014}.
Instead of training a single policy, $m$ different polices are trained to produce diverse behaviors.
The exploration algorithm then chooses from those actions which have not been learned to perform bad, i.e., having high regret, in the current context. 
Again, $m$ is a parameter to be chosen by the user.

Contextual bandits are related to Reinforcement Learning (RL) \cite{Sutton2018}.
The main distinction between bandits and RL agents is that bandits perceive each iteration as independent of the previous one \cite{Lattimore2019}, i.e., the selected action does not affect the next context that is observed. 
In our scenario, the next test case is independent of the test transformation chosen for the previous test case.
RL agents, however, are designed to operate over multiple subsequent iterations, where a chosen action influences the context in the next iteration.
General RL agents could be applied by reducing the length of each scenario to one step, but our early experiments found contextual bandits to be more efficient.

Bandit algorithms have been successfully applied in a variety of domains, such as news article recommendation \cite{Li2010}, advertisement selection \cite{Lu2010,Tang2013}, statistical software testing \cite{Baskiotis2006}, constraint optimization \cite{Loth2013,Balafrej2015}, or real-time strategy games \cite{Ontanon2017}.
In this work, we apply contextual bandits to the selection and configuration of metamorphic relations in software testing.

\subsection{Related Work} \label{sec:relwork}



Metamorphic Testing (MT) has been applied to a variety of domains and applications, see \cite{Segura2016,Chen2018} for an in-depth overview.
Successful application domains include testing of driverless cars \cite{Zhou2019}, search engines \cite{Zhou2016b}, machine translation systems \cite{Sun2018}, performance testing \cite{Segura2018,Johnston2019}, constraint solvers \cite{Akgun2018} or bioinformatics \cite{Shahri2019}.
Previous works already focused on its automation, for example by exploring algorithms to specifically identify fault-revealing inputs \cite{Gotlieb2003} or performing an empirical study on selecting good Metamorphic Relations (MRs) \cite{Mayer2006}.
Other works predict the applicability of a MR for a system \cite{Kanewala2013,Kanewala2016}.
Based on source code traces, a classification model predicts which MRs of a given set can be applied.

Due to the emergent success and usage of machine learning in different application areas, the verification and validation of these systems has received increasing attention.
Several works approach testing machine learning systems based on software testing techniques, such as differential, multi-implementation \cite{Pei2017} or mutation testing \cite{Ma2018}.

Because testing machine learning systems, due to their stochastic nature, is affected by the oracle problem \cite{Barr2015}, there has been work to especially apply MT for this purpose.
Murphy \etal identify a set of general MRs, that hold for a variety of machine learning algorithms \cite{Murphy2008}, and are shown to be effective \cite{Xie2011}.
Chan \etal further use MT to identify violations of MRs from passed test cases of classification models \cite{Chan2010}.
It has also been shown that MT can be used for deep learning-based applications, e.g., to test the classification of biological cells \cite{Ding2017}.
Dwarakanath \etal identify implementation faults in image classifiers \cite{Dwarakanath2018}.
They introduce MRs that affect the training and test data used during model training and demonstrate how these MRs can be applied to find implementation errors in training procedures and model architecture.
Yang \etal propose to test unsupervised clustering methods \cite{Yang2019} and Mekala \etal explore the application of MRs to detect adversarial examples for deep learning models \cite{Mekala2019}.
However, Saha and Kanewala \cite{Saha2019} recently evaluated the effectiveness of MRs for testing supervised classifiers based on mutations of the system under test. 
They found that the detection rates for the used MRs of previous studies are limited when generating a large set of mutants.

Previous work also considered the adaptive control of software testing through feedback while testing \cite{Cai2007,Zhou2018}.
Similar to our method, these works exploit the behavior of the system during the test execution and adjust the testing strategy when the understanding of the software changes.
In these works, the adjustment of the testing strategy focuses on test case prioritization and selection, whereas our method focuses on the generation of follow-up test cases using metamorphic testing and under consideration of many repeated testing cycles.

\section{Adaptive Metamorphic Testing}
\label{sec:method}

\subsection{Overview}

\begin{algorithm}[t]
    \caption{Adaptive Metamorphic Testing with Contextual Bandits\label{alg:method}}
    \begin{algorithmic}[1]
    \Statex \textbf{Input:} $M$: set of MRs; $\mathit{SUT}$: system under test $P$; $TS$: set of test cases; Iter: Number of iterations  
    \Statex \textbf{Output:} $\B$: trained bandit
    
    \State $i\gets 0$
    \State $\B\gets \text{Load \textit{existing} or initialize \textit{new} Bandit}$
    
    \While{$i<\text{Iter}$}
        \State \textbf{Select} $\stc \in TS$  \Comment{Draw source test case from test suite}
        \State $c \gets \B.\text{ExtractContextFeatures}(\stc)$ \Comment{Generate feature vector for source test case $\stc$}
        \State $R_{\phi}\gets \B.\text{SelectBanditArm}(c, M)$ \Comment{Select one MR $R_{\phi}$ using the bandit}
        \State $v\gets \text{Apply}(\mathit{SUT}, R_{\phi}(\futc))$  \Comment{Execute SUT with transformed test $\futc$, get a verdict $v$}
        \State $\B\gets \B.\text{UpdateBandit}(R_{\phi}, c, v)$ \Comment{Train the bandit with the feedback}
        \State $i\gets i+1$
    \EndWhile
    \State \Return $\B$ \Comment{Return updated bandit for future test cycle}
    \end{algorithmic}
\end{algorithm}

In this section, we introduce Adaptive Metamorphic Testing (AMT) with contextual bandits. 
Our method is based on a test transformation bandit that learns to select follow-up test cases from a set of applicable metamorphic test cases.
Algorithm~\ref{alg:method} shows an overview of the main steps of AMT.
At the core of AMT, a contextual bandit receives a description of the source test case.
Based on this context vector, the bandit selects an action, which resembles a MR, and the configuration of this transformation.
Both are applied to generate a follow-up test case.
After generating the follow-up test case, the system under test is executed and the test result evaluated according to the MR acceptance criterion.
The method can be directly deployed without any pre-training step. However, during the first iterations, MR selection is partly random to gather initial experiences about the different MRs effectiveness and their potential payoffs, when applied to the available source test cases.
After several iterations have been performed, the bandit learns to focus on MRs which are most likely to reveal faults.
Nevertheless, the bandit continues to explore among the MRs, i.e., it sometimes chooses MRs which do not promise the highest payoff.
This is important to adjust to changes in the system under test as well as to gather additional information about the effect of MRs in different contexts.

\begin{definition}[Adaptive Metamorphic Testing]
Let $P$ be an implementation of a target algorithm $f$ with $TS$ being its test suite; let $M$ be a set of metamorphic relations applicable on $TS$ and let $\B$ be a contextual bandit.
Adaptive Metamorphic Testing (AMT) is an iterative variant of metamorphic testing and involves the following steps at each iteration:
\begin{enumerate}
    \item A test case $\stc$ is (randomly) selected from the test suite $TS$ and executed to obtain its output $P(\stc)$.
    \item The bandit $\B$ selects a MR $R$ based on the context features of $\stc$.
    \item Construct one or more follow-up test cases $\langle \futc_1, \futc_2, \dots, \futc_k \rangle$ according to $R$ and obtain their respective outputs $\langle P(\futc_1), P(\futc_2), \dots, P(\futc_k))$.
    \item Examine the results with reference to $R$. If $R$ is not satisfied, then this MR has revealed that $P$ is faulty.
    \item Report the results of the execution back to $\B$ for adaptation of the learning algorithm.
\end{enumerate}
\end{definition}

The arm selection of the bandit, i.e., the selection of a metamorphic relation $R$ with its parameters is handled by a hierarchy of contextual bandits using the context features $c$ at each iteration of the algorithm.

On the highest level, the main contextual bandit \B selects one MR $R$ from the set of supported MRs.
Afterwards another action-specific contextual bandit $\B_R$ is queried, using the same context information as the main bandit, for the configuration parameter $\phi$: $\B_R(c_i) \rightarrow \phi$.
If the metamorphic relation does not require additional configuration, the second step is skipped.
The MR $R_\phi$ can then be used to generate a follow-up test case for the current iteration.
After test execution, both bandits \B and $\B_R$ are trained from the received feedback.
\method consists of one main contextual bandit plus one additional contextual bandit for each configurable MR.

Adaptive Metamorphic Testing is independent of the application domain or the implementation or else the specific MRs that can be applied.
It only takes as inputs a set of MRs, the system under test, a set of test cases, and a user-defined parameter corresponding to the maximum number of iterations to run. As output, the method returns a trained bandit $\B$ which has learned to select the MRs which have the greatest chances to detect faults in the system under test.  

\subsection{Components}
Adaptive Metamorphic Testing requires only a few system-specific components.
In the following, we discuss each of these components.

\subsubsection{Extract Context Features}

In order to select an appropriate MR  which is likely to reveal a fault, for a source test case, it is mandatory to feed the contextual bandit with relevant context information about the source test case.
This context is captured with a feature vector, which is a real-valued vector of fixed size $n$.
The function receives the source test case as input and returns the feature vector: $ExtractContextFeatures: \mathcal{S} \rightarrow \mathbb{R}^n$.

By using representative features, the contextual bandit can learn a mapping from the source test cases, which are described through the features, to the MRs.
For that, it has to include test characteristics that can be affected by the metamorphic relations.
Therefore, the features need to capture the necessary details distinctive about the individual test case, especially those that relate to the effect of the MR.

How the feature vector is formed, is a domain-specific problem and requires some degree of domain knowledge.
For example, when testing scientific software for matrix calculations \cite{Mayer2006}, the features should describe the characteristics of the original input matrix in order to select which transformation is applied.
In our experiments, which are based on testing computer vision problems, we rely on common feature modeling used in machine learning. For instance, using a pre-trained neural network to extract image features is common in computer vision.
A similar approach could be used for text processing, where textual features, e.g., used vocabulary, text sentiment, or sentence structure, can be derived using pre-trained networks.

\subsubsection{Metamorphic Relations}

The most relevant component to acquire for applying Adaptive Metamorphic Testing is the set of MRs, which is highly domain-dependent.
The automatic and systematic identification of MRs for a system is an ongoing research topic \cite{Chen2018}, but in many cases, MRs can be extracted by using domain knowledge about the system or reviewing the existing literature from the Software Testing community.
In the special case of testing machine learning systems, a starting point to uncover MRs is to exploit data augmentation methods that are commonly used.
As shown in our experiments, these augmentations can serve as a basis for MRs and help to identify weaknesses in ML systems.

\subsubsection{Select Bandit Arm}
The selection of an appropriate MR is mostly handled by the internal contextual bandit algorithm and does not have to be individually implemented for a new system under test.
The main bandit selects the MR and, if necessary, the action-specific bandit for this MR selects the parameter to configure the MR.
Nevertheless, the configuration of the contextual bandit influences the performance of the system and should be adjusted, depending on the number of available MRs and the robustness of the SUT.
This allows us to focus on exploiting MR that reveal faults in the system or to broadly explore the effects of many different MRs.

The most important configuration parameter to adjust is the exploration rate, i.e., how often does the bandit choose a different action than the most promising one.
Using a high exploration rate allows us to examine different combinations of test cases and systems, which is relevant to detect new faults in the system and extend the coverage of different tests. 
Conversely, a lower exploration rate exploits combinations of tests and MRs that have often fail previously.
Traditionally, Metamorphic Testing often creates follow-up test cases at random, which corresponds to a maximal exploration rate here.

Exploitation is relevant when repeatedly testing the system, e.g., in continuous integration settings, or when trying to understand the weaknesses of the system for a certain group of MRs.
Still, it is not only desired to exploit known weaknesses, but broad coverage of the system behavior is desired for higher test confidence.
If the behavior of the system changes, because it is becoming more robust to previously effective MRs, the bandit learns this and can adjust its selection for future iterations.
Conclusively, compared to other applications of contextual bandits, where especially the exploitation of known good actions is in focus, exploration is more prominent in Adaptive Metamorphic Testing to broadly test the system behavior.

\subsubsection{Transform, Execute and Evaluate}
The selected MR and its configuration transform the source test case into a follow-up test case.
That test case is then executed and the test verdict is evaluated according to the MR.
These core steps of Adaptive Metamorphic Testing are similar to those required in traditional MT.

\subsubsection{Update Bandit}
After the follow-up test case has been executed and the results have been evaluated, the bandit's policy is updated.
This requires information about the initial context feature vector, the chosen MR and its configuration, and the test verdict.
The {\it update routine} updates the expected reward for this MR.
The exact update routine is specific to the contextual bandit algorithm and its configuration and we refer to the corresponding literature for its description \cite{Langford2007,Lattimore2019}.

Nevertheless, choosing the appropriate reward for a failed test case has to be done while adjusting the bandit for a new system to test.
In most scenarios, where the goal is to find the most fault-revealing MRs, as described below in Section~\ref{sec:faultrevealing}, the reward is the same for every failing test case.
However, if the bandit has the goal to identify certain properties of the SUT, it can be necessary to propose a different reward structure that depends on the selected MR.
This second scenario is further described in Section~\ref{sec:experiment_mrs}.




\section{Application Scenarios of Adaptive Metamorphic Testing} \label{sec:applications}

Contextual bandits are powerful to explore the effects of the MRs in different contexts, but can also exploit the gathered experiences to subsequently focus on those relations that are most likely to reveal faults.
From these properties, we identify two application scenarios of Adaptive Metamorphic Testing that we discuss further and evaluate as part of the case studies.

\subsection{Fault-Revealing MR Selection} \label{sec:faultrevealing}
The first application of Adaptive Metamorphic Testing is the selection of metamorphic relations which are prone to reveal faults in the system under test. 
In cases where the MR has parameters, an additional contextual bandit is responsible to select these parameters, as described before.
This application, which we refer to as \textit{fault-revealing MR selection}, steers MT towards greater effectiveness when there are many MRs available and not sufficient resources to apply them all. In this application, all MRs are considered distinct from each other. 
Accordingly, the achievable reward received for revealing a fault is identical for all MRs.

\subsection{Robustness Boundaries} \label{sec:robustness}
The second application uses the exploration/exploitation trade-off of contextual bandits to identify \emph{robustness boundaries} of the software under test.
With robust boundaries, we focus on MRs whose effect can be adjusted by user-defined parameters, especially those with continuous or a range of discrete values that control the distance between source and follow-up test cases.
As an example taken from the case studies, while testing an image analysis system, one possible transformation is to rotate the image, where the degree of rotation is a user-defined parameter.
If the system is susceptible to treat wrongly rotated images, it is likely that large rotations, e.g. by 90 degrees, are more likely to cause mistakes than smaller rotations.
By identifying the robust boundaries, information can be inferred about the trade-off between acceptable transformations and exceedingly strong manipulations.
This result yields both a robustness characteristic and a starting point for a more curated set of requirements on the system under test.

\section{Experimental Evaluation} \label{sec:experiments}

\subsection{Case Studies}
We consider two case studies to evaluate our tool \method.
Both case studies come from the field of digital image processing, where deep learning methods commonly represent the state-of-the-art approaches~\cite{Pouyanfar2018}: \emph{image classification} and \emph{object detection}.
For each of the case studies, we consider both previously introduced application scenarios (see Section~\ref{sec:applications}) and identify fault-revealing MRs as well as robustness boundaries against configurable image transformations.
We have formulated the following three research questions as a guideline for our experiments:
\begin{itemize}
    \item[\textbf{RQ1}] Does Adaptive Metamorphic Testing, implemented as \method, learn to select MRs whose follow-up test cases reveal faults?
    \item[\textbf{RQ2}] Is AMT effective to approximate the distribution of faults in the system under test?
    \item[\textbf{RQ3}] Is AMT computation- and data-efficient compared to random sampling of MRs and exhaustive search of all follow-up test cases?
\end{itemize}

Previous work for testing image processing applications has considered random and metamorphic testing \cite{Mayer2006a,Guderlei2007}, but focused on the evaluation of handcrafted image processing applications, whereas we focus in our experiments especially on machine learning-based computer vision systems.
The existing studies focus on testing the basic functionality of the system by generating random images and transforming them using a set of transformations, different from our approach where we base on an existing dataset of images from the domain that the ML model has been trained on. 
We furthermore especially consider the selection of good MRs.
Xu \etal recently presented another use case for metamorphic relations in image classification applications beyond testing \cite{Xu2018}. 
Their work uses metamorphic relations, based on separation and occlusion, to augment the training data and fine-tune the model.

In the following, we present the two case studies, image classification and object detection, with their setup and the considered MRs.
We further discuss the configuration of \method, and finally, present the experimental results and our findings.

\subsection{Image Processing Applications} \label{sec:casestudies}

\begin{figure}
    \centering
  \begin{subfigure}[t]{0.49\textwidth}
    \centering
       \includegraphics[width=0.725\textwidth]{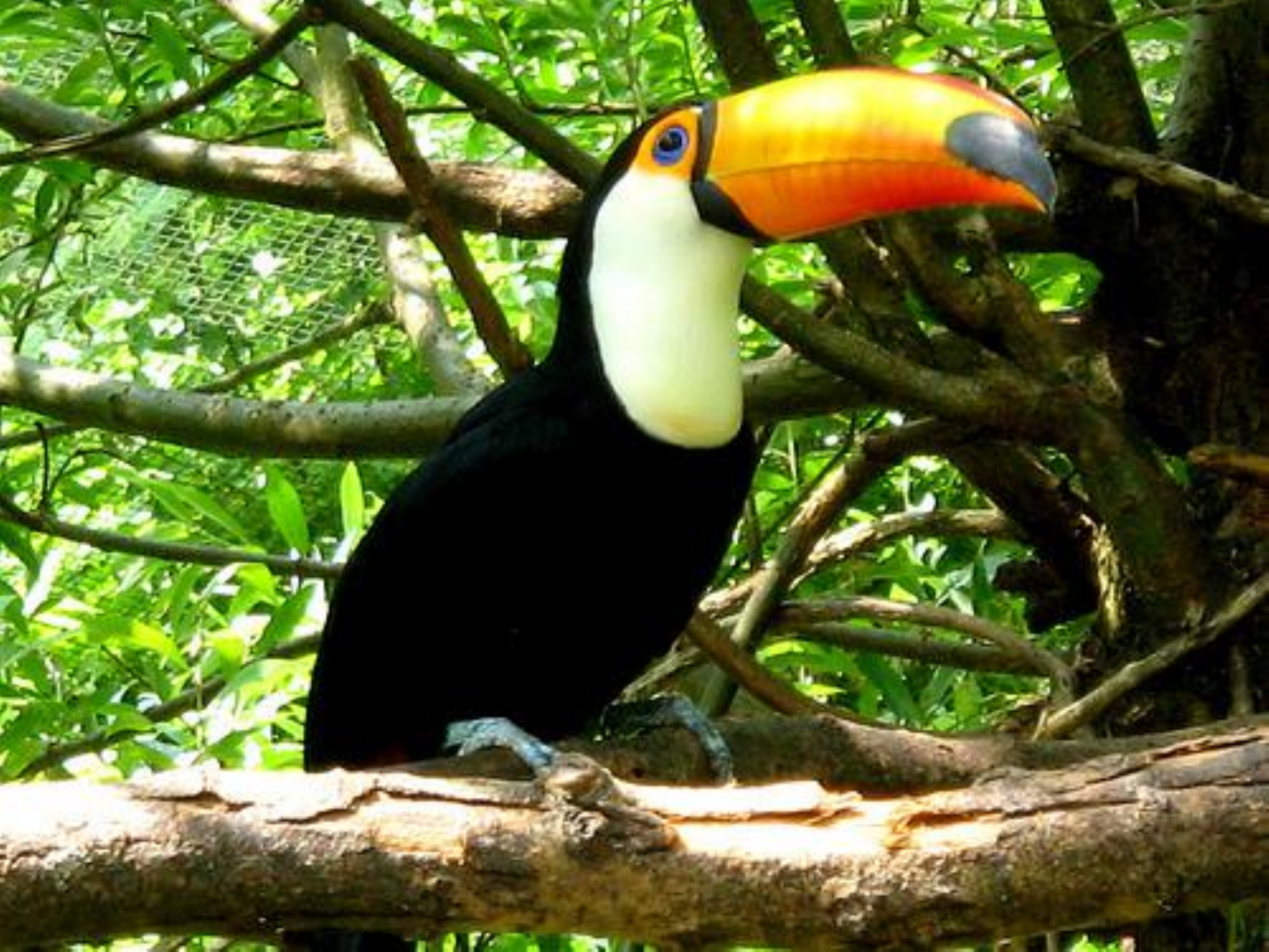}
       \caption{Image Classification Example (from ImageNet~\cite{Russakovsky2015}). The goal is to
         assign a single class label to the image, e.g. \emph{toucan} here.}
  \end{subfigure}\hfill
  \begin{subfigure}[t]{0.49\textwidth}
    \centering
    \includegraphics[width=0.75\textwidth]{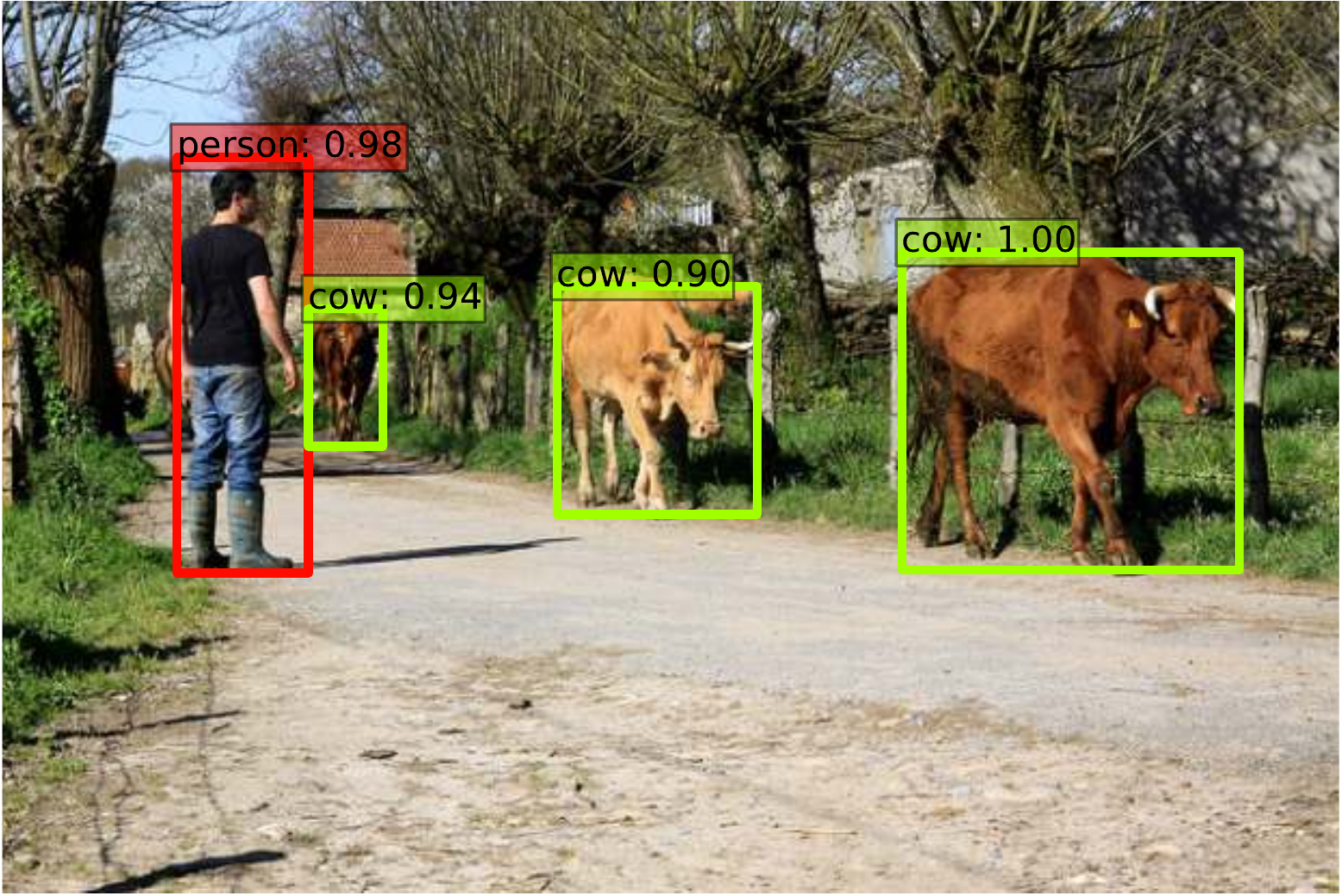}
    \caption{Object Detection Example (from \cite{Liu2016a}). The goal is to identify and categorize objects by drawing a bounding box and assigning a class label.}
    \end{subfigure}
    \caption{Image classification and object detection examples.}
    \label{fig:images}
\end{figure}

We describe two case studies where testing image processing systems is necessary (see \figurename~\ref{fig:images}).
In the first case study, an image classification system is tested. 
The second case study focuses on an object detection application.

\subsubsection{Image Classification} \label{sec:classification}

An image classification task, also image recognition, has the goal to identify the object shown in an image, e.g., assign the image to one of a fixed number of classes.
Since 2012, the state-of-the-art method for image classification, among other image analysis tasks, are deep neural networks, such as residual neural networks (ResNets)~\cite{He2016} or SqueezeNet~\cite{Iandola2016}.

In our case study, we test a SqueezeNet model that has been initially trained on the ImageNet dataset~\cite{Russakovsky2015} and then fine-tuned for the 10 classes of the CIFAR-10 dataset~\cite{Krizhevsky2014a}.
For testing the model, we use the CIFAR-10 test set, consisting of $10,000$ labeled images.

Following the metamorphic relation between source and follow-up test cases, we consider a test as failed, if the transformed image leads to a different classification result than the original image.
The correctness of the original class prediction does not influence the test result, because the bandit does not know the initial model performance.
Instead, the bandit aims to select transformations that affect the outcome in a fault-revealing manner compared to the original output.
Testing the difference in outputs between source and follow-up test cases removes the dependency from having to use labeled data, i.e. data where the ground-truth class is known and allows the integration of other data sources.
Nevertheless, for testing the system, we monitor also the accuracy of the system for correctly classifying the images, but we do not use this information as feedback for the test transformation bandit, although that would also be a viable setup.

\subsubsection{Object Detection} \label{sec:detection}

Object detection is a generalization of the image classification task in the sense that there can be multiple objects on a single image. 
Besides assigning classes to these objects, it is also necessary to provide bounding boxes around the location of each object.
This means that the output of an object detection model consists of a class label and four coordinates for the bounding box for each detected object.
Object detection systems employ deep neural networks of a similar, but extended, architecture compared to image classification systems.

The system to test in this case study is a pre-trained object detection model, based on the open-source TensorFlow Object Detection API\footnote{TensorFlow Object Detection API: \url{github.com/tensorflow/models/tree/master/research/object_detection}} \cite{Huang2017b}.
In particular, we test an implementation of a single shot multibox detector (SSD) \cite{Liu2016a} with a feature pyramid network (FPN) \cite{Lin2017b}, based on a ResNet-50 network \cite{He2016}.
We refer the reader to the given references for an in-depth overview of the models. We see the system to test here as a black box.
However, briefly said, the model detects objects in images with a single neural network by assigning one of multiple predefined box sizes, their size adjustment and classification scores at the same time.
By reducing the complexity to a single neural network, it is a fast model for real-time object detection that achieves state-of-the-art performance.
The used model was trained on Microsoft COCO dataset \cite{Lin2014} and is available within the Object Detection API\footnote{The exact name in the object detection model zoo is \textit{ssd\_resnet50\_v1\_fpn\_shared\_box\_predictor\_640x640\_coco14\_sync\_2018\_07\_03}}.

For testing, we use $5,000$ images from the validation set of the \textit{MS COCO challenge 2017} as source test cases.
We apply the same input transformations on the images as in the image classification case study and as described in Section \ref{sec:experiment_mrs}.
Because each image annotation consists of an additional bounding box per object in the image, the metamorphic relations are extended to transformation also on the bounding boxes.
For example, rotating the image rotates the bounding box of the object to match the rotated object, and flipping the image from left to right also flips the positions of the bounding boxes in the image (see the next Section~\ref{sec:experiment_mrs} for an introduction of the applied transformations).

Furthermore, we consider the different evaluation metrics for object detection tasks.
In image classification, the result is easily verified by comparing the estimated class with the ground truth class.
In object detection, it is necessary to evaluate the overlapping regions between the estimated bounding boxes and the ground truth, which is called the intersection-over-union (IoU), in addition to the class label of each box:
$$IoU(A, B) = \frac{A \cap B}{A \cup B}$$
where $A$ are the proposed pixels from the object detection model and $B$ is the ground-truth from the dataset.
If the IoU value exceeds a certain threshold, the object is counted as correctly detected.
We follow the evaluation guidelines from the MS COCO challenge and compare the results for the mean average precision (mAP) which is calculated over all objects in an image and the average of different IoU thresholds $0.5, 0.55, 0.6, \dots, 0.95$, for both the original and the transformed image.
If the mAP of the transformed image is below the original mAP minus a performance reduction of 0.05 which is close to 10\% of the average model performance, we interpret the test case to be violating the MR and therefore as failed.

\subsection{Metamorphic Relations and Rewards} \label{sec:experiment_mrs}

In the case studies of this paper, we use tests where the input is an image and the output is a classification of this image to recognize certain objects.
The considered MRs are all related to image transformations, e.g. mirroring or rotating, under the property that the results of image classification and object detection do not change, i.e., the MR defines equality of the outputs, while the input is transformed.

In most cases, these transformations must only not modify the class of these images.
However, in object detection, some transformations of images that impact object location markers entail similar transformation over the outputs.
Here, the MR defines a relation between the outputs that is similar to the transformation of the input.

We select seven common image transformations among all possible transformations as MRs, some of which have been used in previous work on metamorphic testing for image analysis methods \cite{Mayer2006a,Mekala2019}.
Among these MRs, two are configurable by an additional parameter $\phi$. 
These transformations are the following: 
\begin{enumerate}
\item Blur the image by the average value of neighbor pixels (Blur)
\item Flip the image from the left to the right (Flip L/R)
\item Flip the image upside down (Flip U/D)
\item Convert a colored image to grayscale (Grayscale)
\item Invert the colors of the image (Invert)
\item Rotate the image by $x$ degrees (Rotation)
\item Shear the image by $x$ degrees (Shear)
\end{enumerate}

\begin{figure}[t]
  \centering
  \begin{subfigure}[t]{0.19\textwidth}
    \includegraphics[width=\textwidth]{figures/4800_original.pdf}
    \caption{Original Image}
  \end{subfigure}\hfill
  \begin{subfigure}[t]{0.19\textwidth}
    \includegraphics[width=\textwidth]{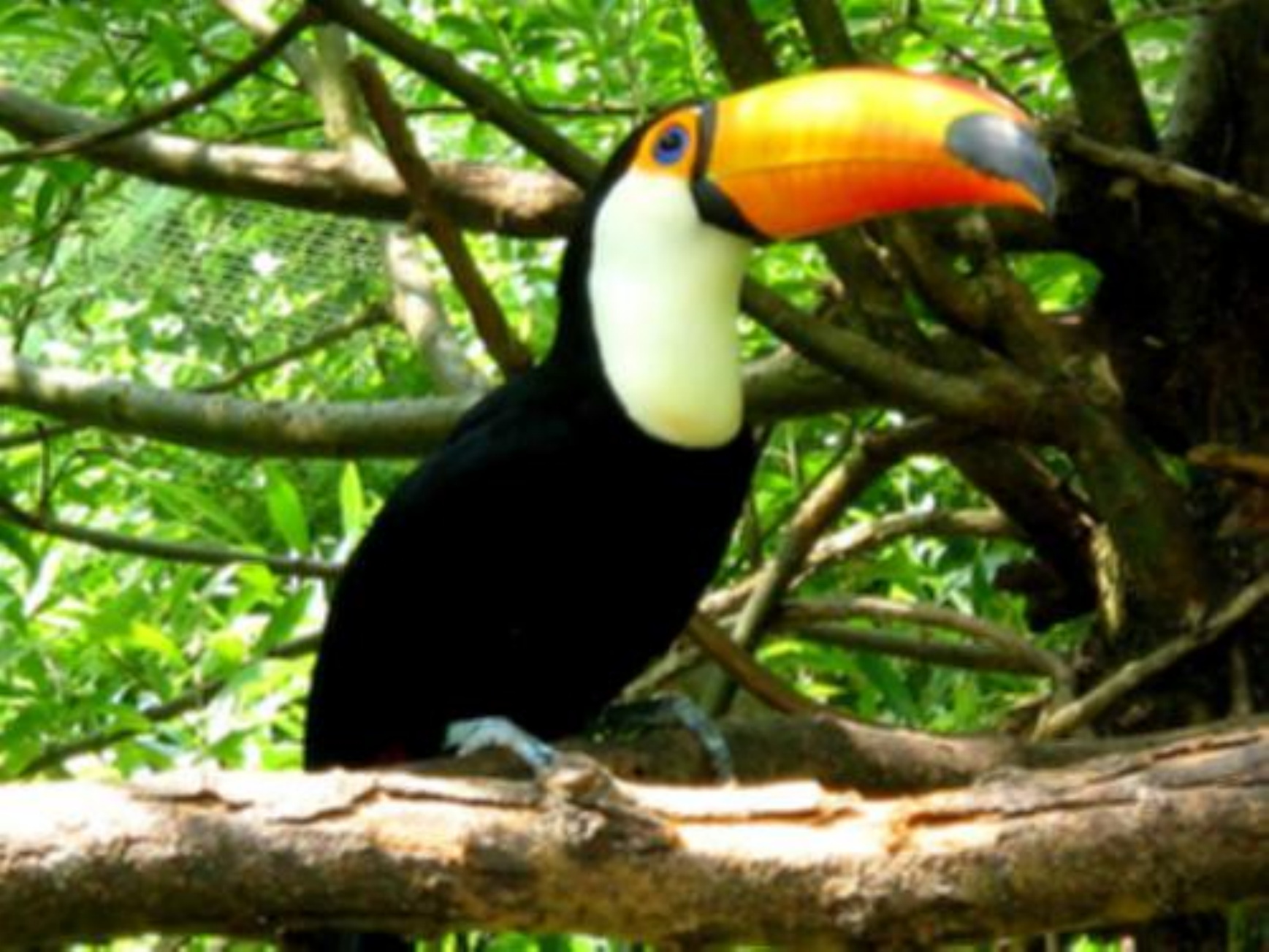}
    \caption{Blur}
  \end{subfigure}\hfill
  \begin{subfigure}[t]{0.19\textwidth}
    \includegraphics[width=\textwidth]{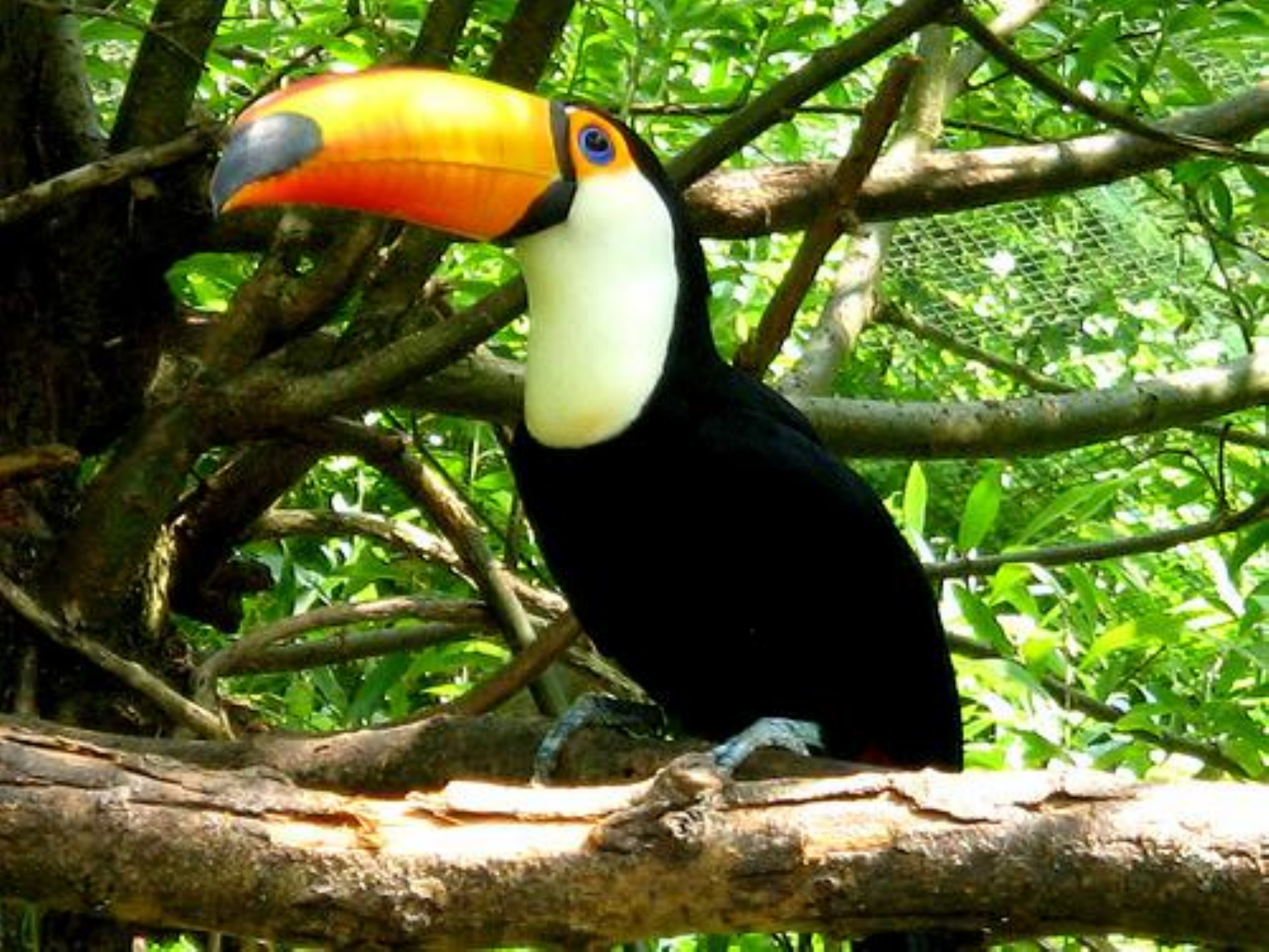}
    \caption{Flip left $\leftrightarrow$ right}
  \end{subfigure}\hfill
  \begin{subfigure}[t]{0.19\textwidth}
    \includegraphics[width=\textwidth]{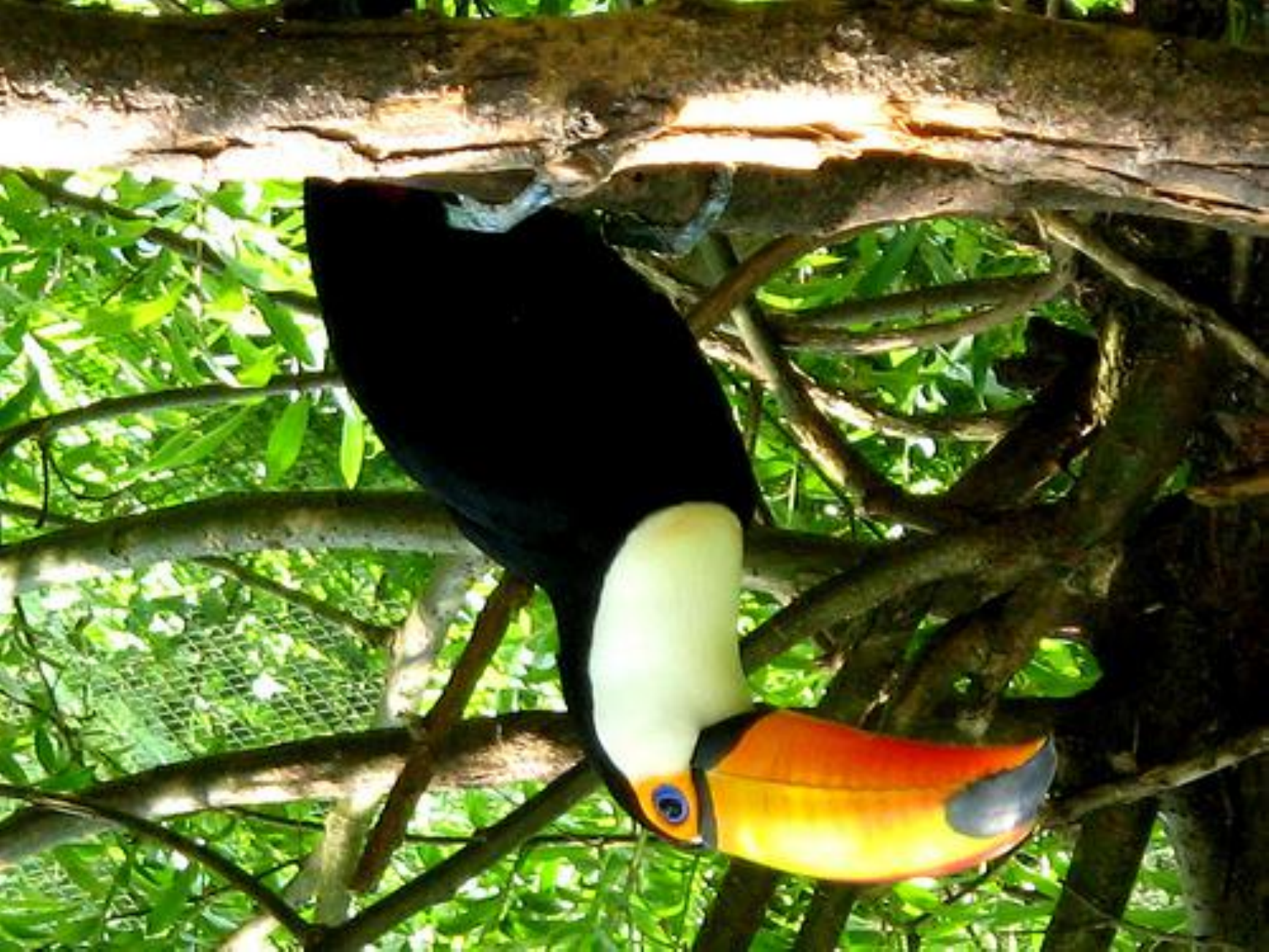}
    \caption{Flip up $\leftrightarrow$ down}
  \end{subfigure}\hfill
  \begin{subfigure}[t]{0.19\textwidth}
    \includegraphics[width=\textwidth]{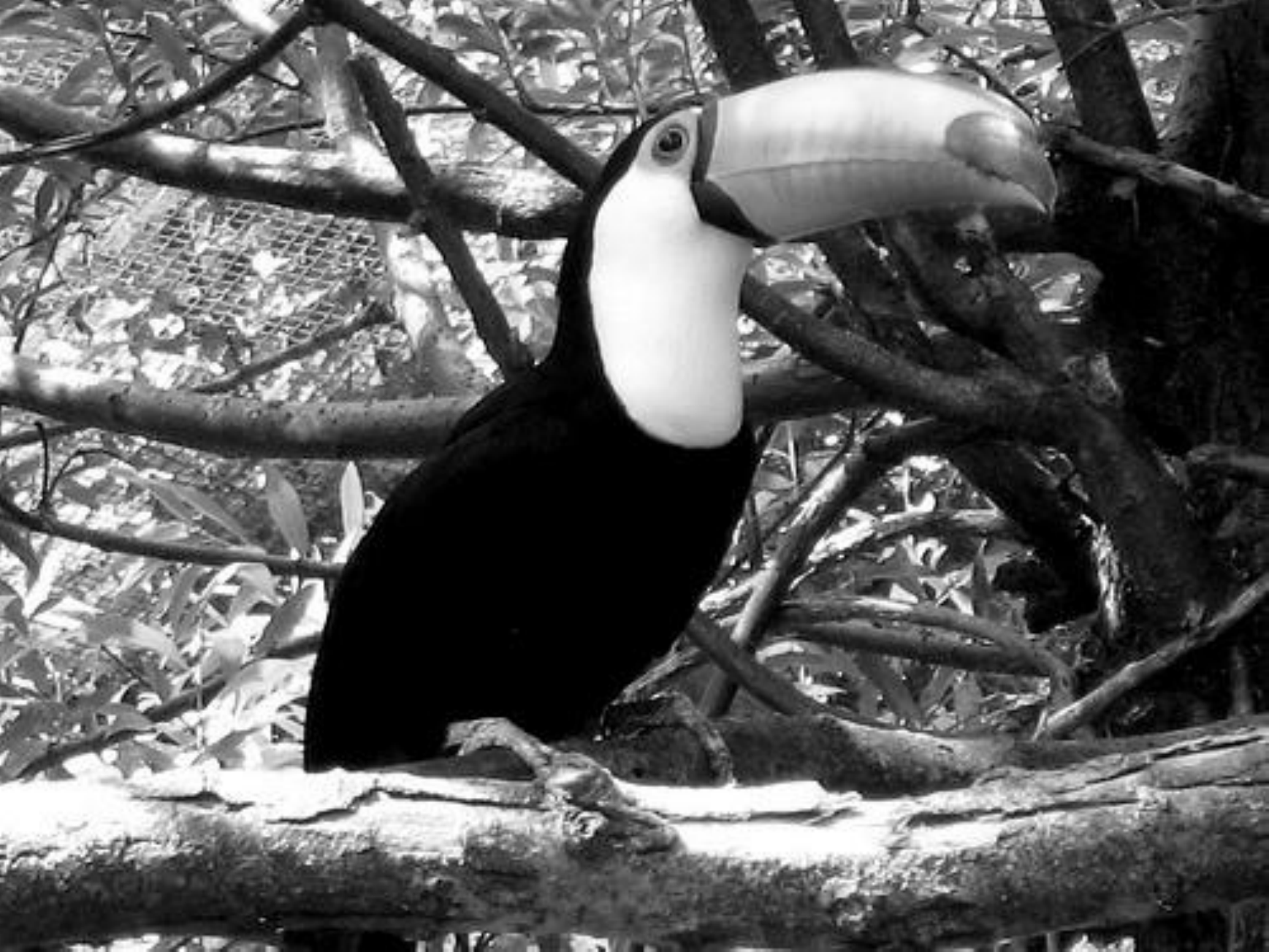}
    \caption{Convert to grayscale}
  \end{subfigure}\\

  \begin{subfigure}[t]{0.19\textwidth}
    \includegraphics[width=\textwidth]{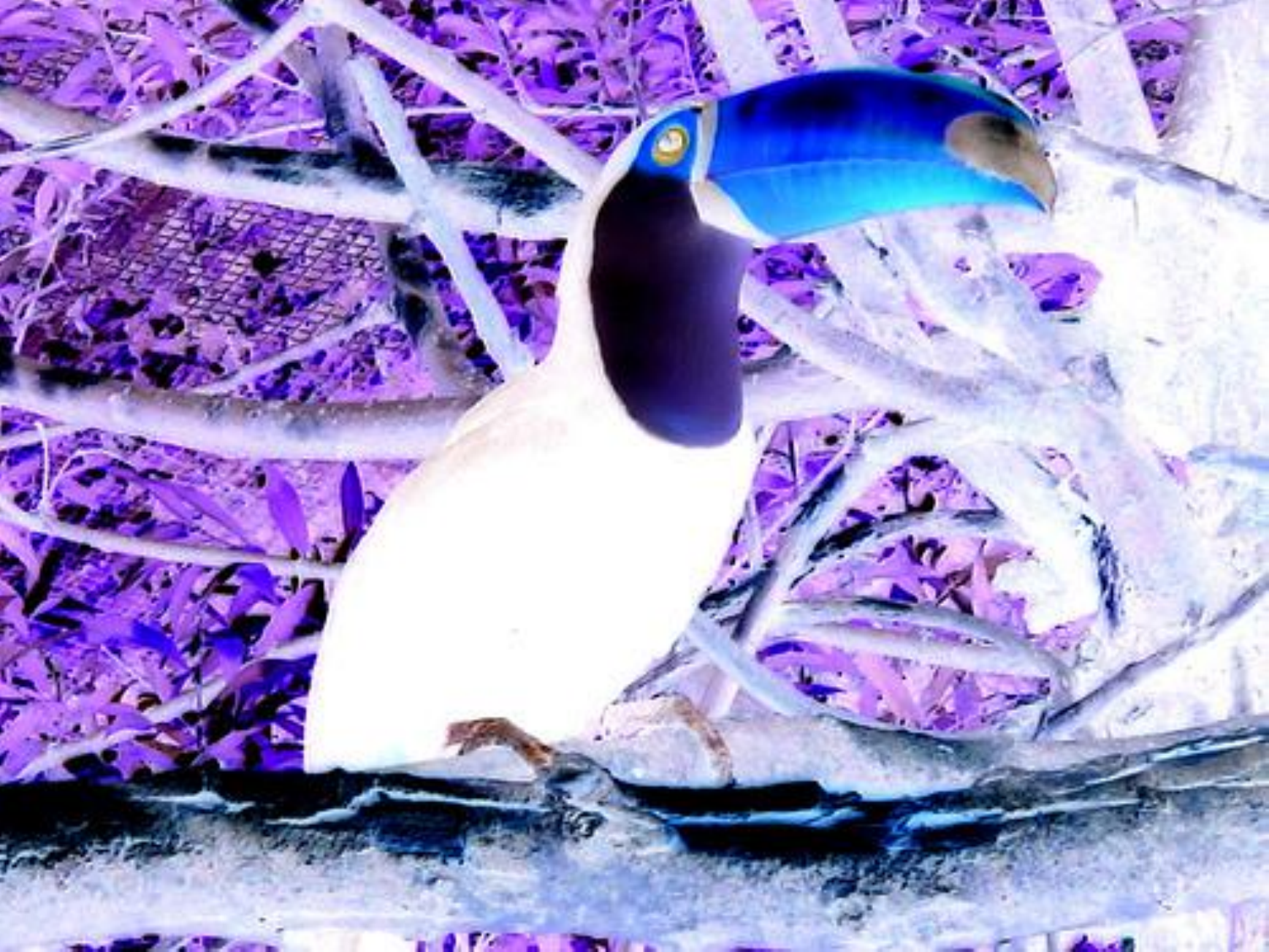}
    \caption{Invert the image}
  \end{subfigure}\hfill
  \begin{subfigure}[t]{0.19\textwidth}
    \includegraphics[width=\textwidth]{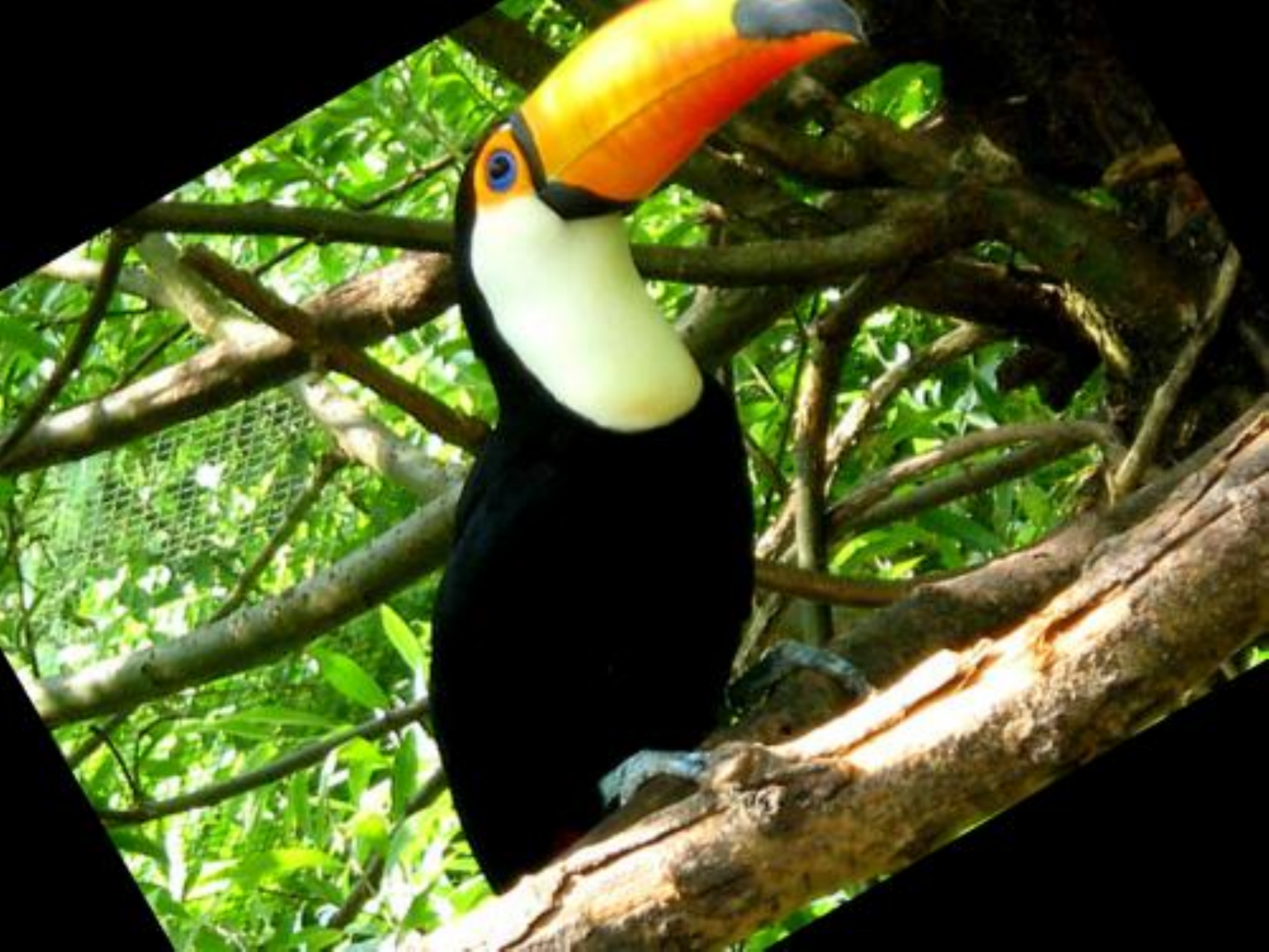}
    \caption{Rotate -30 degrees}
  \end{subfigure}\hfill
  \begin{subfigure}[t]{0.19\textwidth}
    \includegraphics[width=\textwidth]{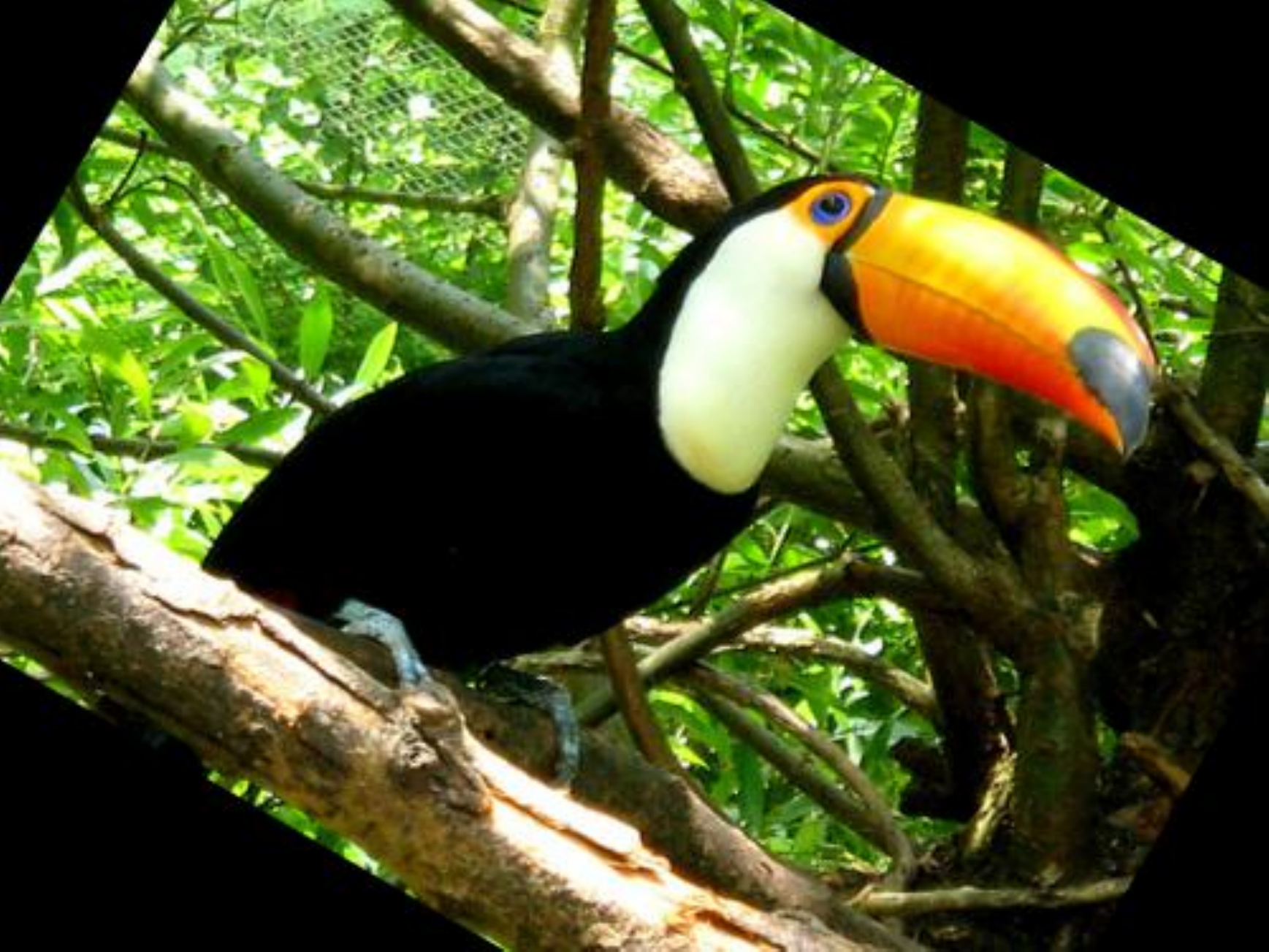}
    \caption{Rotate $+30$ degrees}
  \end{subfigure}\hfill
  \begin{subfigure}[t]{0.19\textwidth}
    \includegraphics[width=\textwidth]{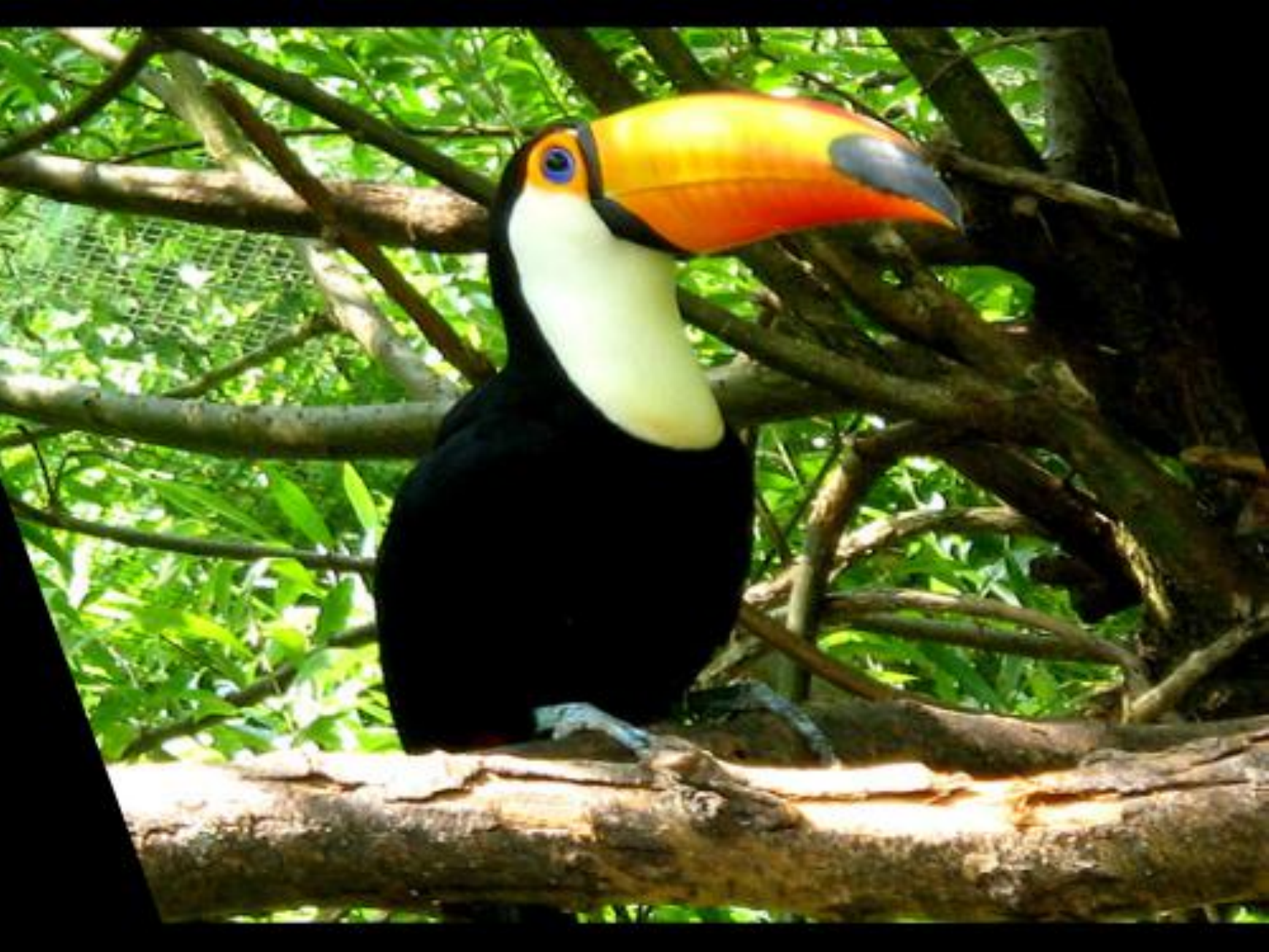}
    \caption{Shear -20 degrees}
  \end{subfigure}\hfill
  \begin{subfigure}[t]{0.19\textwidth}
    \includegraphics[width=\textwidth]{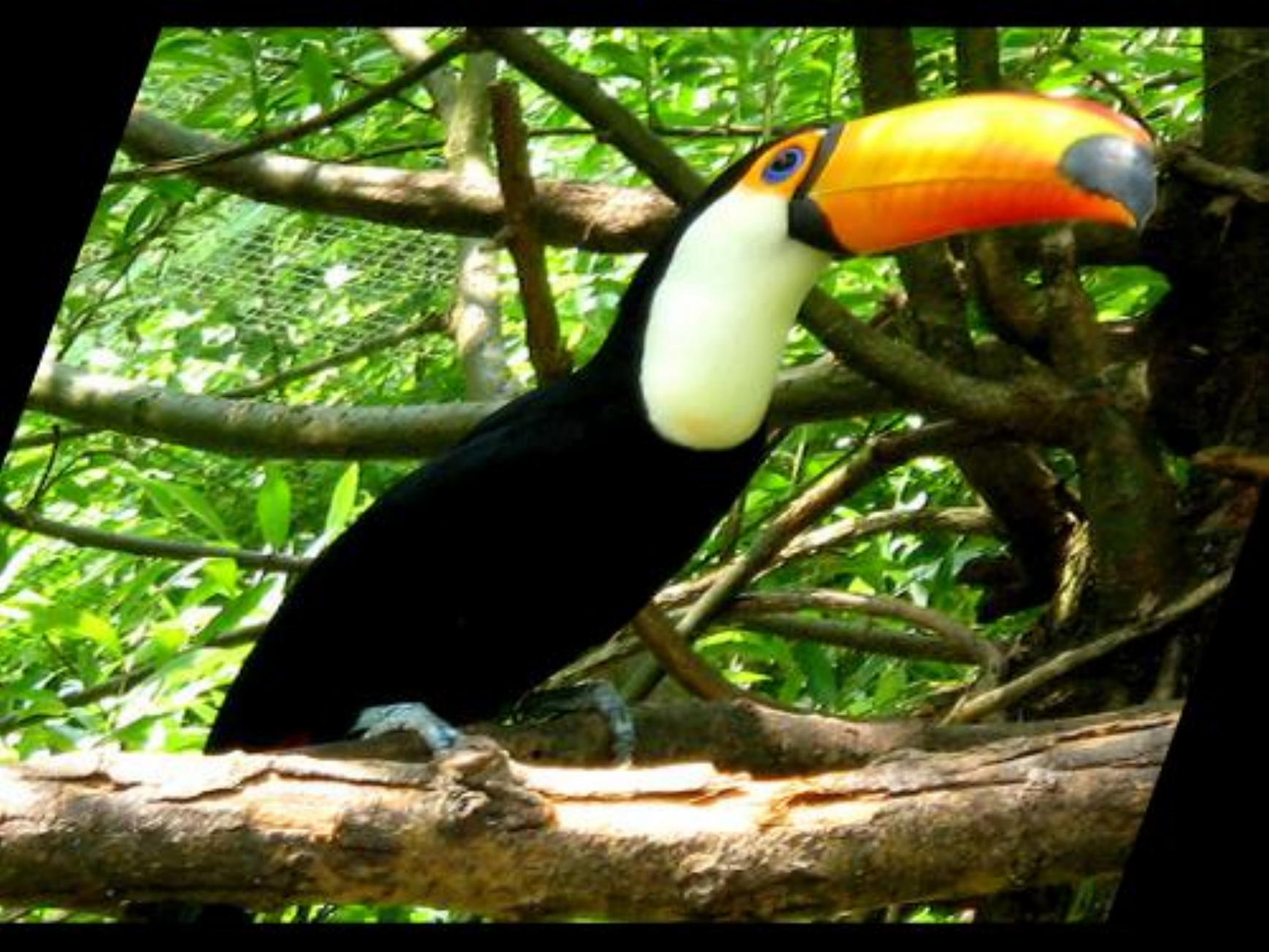}
    \caption{Shear +20 degrees}
  \end{subfigure}
  \caption{Metamorphic relations, i.e. image transformations, and their effects. Image taken from the ImageNet dataset~\cite{Russakovsky2015}.}
  \label{fig:mr_effects}
\end{figure}

The effects of these transformations are shown in Figure~\ref{fig:mr_effects}.
The MRs Rotation and Shear expect a parameter to define the transformation effect. 
For Rotation, we consider $36$ distinct values in steps
of 5 degrees in the degree range $[-90;90]$, excluding rotations of $0$ degrees.
For shear, we include $18$ values in the range $[-45;45]$ in steps of $5$ degrees, again excluding $0$ degrees.

The reward for the main MRs is set up such that a failed test case is rewarded with $1$ and a passed test case with $0$, independent of which MR was selected.
For the action-specific bandits, which select the MR parameters for Rotation and Shear, the reward structure is designed to encourage the selection of the smallest failing parameters. 
Therefore, the smallest parameter of $-5$ respectively $5$
degrees receives a reward of $10000$, if revealing a fault. 
For each additional step, the reward is divided by two. 
Thereby, choosing a smaller parameter value has always higher payoff than a larger rotation, if successful.

\subsection{Implementation}
\begin{figure}
    \centering
    \includegraphics[width=\textwidth]{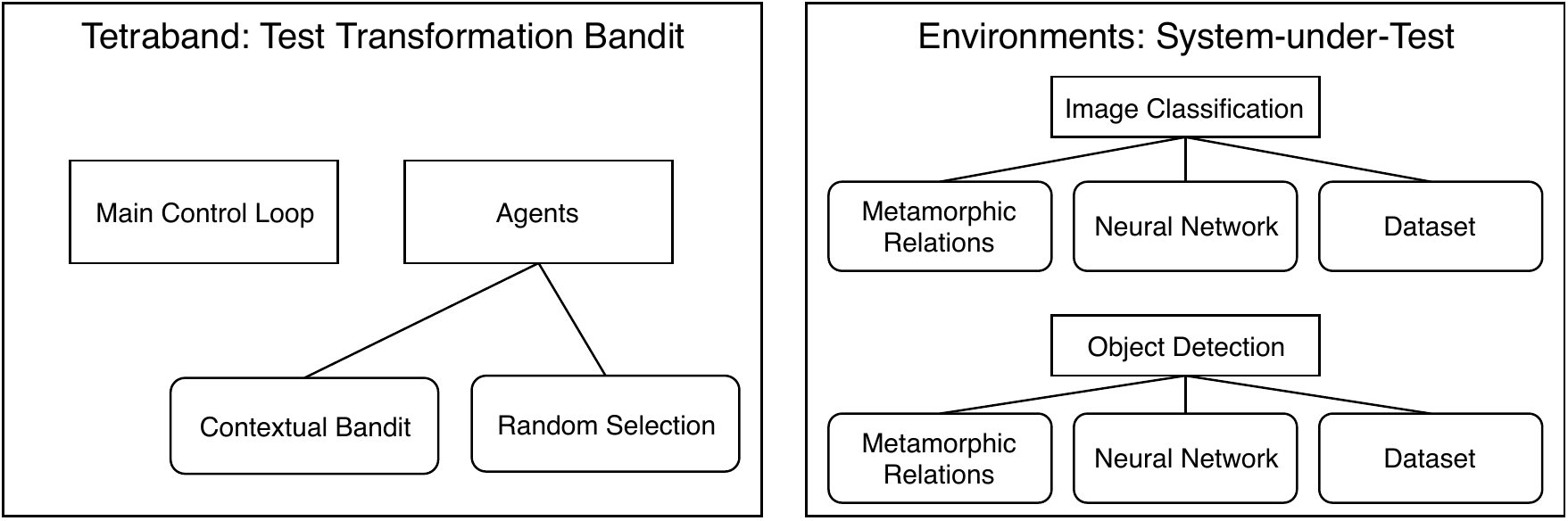}
    \caption{Overview of \method, our implementation of Adaptive Metamorphic Testing. It is a light-weight implementation that allows extension and adaptation to other environments and settings.}
    \label{fig:architecture}
\end{figure}

We implemented Adaptive Metamorphic Testing in a tool called \method.
Our implementation package is available at \url{http://github.com/helges/tetraband}.
The software is implemented in Python 2.7 and it is structured into two main components, as shown in \figurename~\ref{fig:architecture}.

One component provides the SUTs used in our case studies, i.e. the image classification and object detection systems.
These SUTs are encapsulated via the OpenAI Gym\footnote{OpenAI Gym: \url{https://gym.openai.com/}} interface \cite{Brockman2016}.
Having separate, standardized environments allows easier reproduction of the experiments and their usage in other work.
Their functionality includes the feature extraction for each SUT, as well as the application of the available MRs.
The metamorphic relations for image manipulation are realized with the imgaug library (version 0.2.6) for image augmentation\footnote{imgaug: \url{https://imgaug.readthedocs.io/}}. 
Further details for the setup of the SUTs are given in the description of the case studies in Section~\ref{sec:casestudies}.

The second component is \method itself, our implementation of AMT.
It is mostly an adaptation of a contextual bandit as the main actor, using the machine learning library Vowpal Wabbit 8.6.1\footnote{Vowpal Wabbit: \url{github.com/VowpalWabbit/vowpal_wabbit}}.
Additionally, for comparison, we include a random agent that uniformly picks an arbitrary MR and configuration.

The contextual bandits use the doubly robust policy evaluation algorithm for learning and action selection \cite{Dud2011}.
Exploration is performed through a combination of epsilon-greedy exploration, where it chooses a random action in 10\,\% of the iterations, and online cover exploration \cite{Agarwal2014} with three policies. 
The policy itself is approximated by a feed-forward artificial neural network with a single hidden layer with 16 neurons.
We choose a moderately high exploration rate, because we not only want the bandits to converge on few single actions that repeatedly provide high payoff, but we also want to learn about the effectiveness of other actions.
An important aspect of contextual bandits is that the exploration is not reduced or disabled after training, but stays active although self-adjusted to a lesser extent than in the initial iterations. 
This helps to re-use the bandit to test the SUT repeatedly, e.g. in Continuous Integration, because it can adapt to changing behaviors in the SUT.

\subsection{Experimental Setup}

At each iteration, the source test case is formed by an input image and its annotations from the data set.
The context feature vector is extracted through an additional neural network \cite{SharifRazavian2014} which processes the image and returns a feature vector of $512$ floating-point numbers. 
The network is a pre-trained ResNet-18 network \cite{He2016} from the PyTorch Torchvision model zoo\footnote{PyTorch: \url{https://pytorch.org} / \url{https://github.com/pytorch/vision}}.
The final layer, that usually outputs the identified image class, has been removed and the output of the previous layer is used as the feature vector.
We have also experimented with perceptual image hashing for feature extraction but did not find the hash value expressive enough.

\subsection{Fault-Revealing Metamorphic Relations}

The first experiment focuses on identifying general weaknesses in the system, i.e. identifying which metamorphic relations are fault-revealing for the system under test.

The bandit can freely choose from the seven transformations described in Section~\ref{sec:experiment_mrs} and their reward structure. 
A follow-up test case fails, i.e. it violates the MR, if the transformed image is classified differently than the source image: $SUT(\stc) \neq SUT(\futc)$.
We define the MR to produce equal outputs for the source and follow-up test cases, but we could instead also consider the annotated ground-truth labels for comparison.
Using the difference in outputs between source and follow-up test cases reduces the dependency on this labeling, and is thereby applicable to unlabeled datasets.

\subsection{Robustness Boundaries}
The second experiment takes two specific transformations and learns the robustness boundaries of the SUT for these transformations.
As the robustness boundary, we describe the parameterization of the transformation which changes the source test case as little as possible but is most likely to reveal a fault.

For evaluation, we use the two parameterized MRs that are already used in the other experiments, i.e. Rotation and Shear transformation. 
For both MRs the same parameter space as described in Section \ref{sec:experiment_mrs} is kept. 
The main difference in this experiment is that the focus lies on a single MR and its parameterization. 
This allows us to specifically examine the weaknesses of the SUT towards one transformation and learn about its robustness.

\section{Experimental Results} \label{sec:results}

For each case study, we have performed two experiments: identifying fault-revealing MRs and robustness boundaries.
Each run consists of one pass over the full training set, i.e. 10,000 iterations for image classification and 5,000 iterations for object detection.
For each of the experiments, we show the mean result of 10 runs with different random seeds.
Our findings underline the effectiveness of \method for controlling the metamorphic testing process in software testing, especially for testing machine learning systems.

\subsection{Effectiveness of Metamorphic Relations on Image Classification}

Before the evaluation of \method, we first analyze the effectiveness of the selected MRs for our experiments on the image classification dataset, i.e. CIFAR-10.
We aim to understand whether there are different effects of different MRs and how they affect certain classes of images.
To this end, all MRs were applied to all images of the dataset, which we refer to in the other experiments as the \emph{baseline} reference and the changes in the predicted classes are observed.
The baseline results reported for image classification correspond thereby to the average violation rate over all classes as shown in the rightmost column of the table.

\tablename~\ref{tab:effects_per_action} shows the percentage of images in a class that is affected by a MR, such that they are wrongly classified afterward.
CIFAR-10 consists of ten classes of images, printed as column names.
The effectiveness of the image transformations varies between both MRs and image classes.
The MR Flip U/D is an example of a particularly effective transformation that affects over $50$\% of the images in the test set.
However, when noting the different classes in the dataset, all images have a clear vertical orientation and the flipped image of the objects is unlikely to occur in the dataset, for example for automobiles where $74.6$\% of the images are misclassified.
Images of frogs or airplanes, where the perspectives of the images vary more naturally are less affected, but still to a moderately high degree compared to other MRs.
This is different for the MR Flip L/R where the horizontal orientation is reversed, which corresponds to a variation that is already included in the training data and therefore is the least effective MR.
Other MRs identify stronger differences between classes.
Converting the image to grayscale has little effect on airplanes or ships, which commonly have few distinct features related to color, but large effects for birds or dogs, where colors are more distinctive characteristics.

Conclusively, from the experiment using the image classification dataset, we see different effects per MR and image, which underlines the motivation to learn which MRs are effective for a particular image.

\begin{table}[t]
    \centering
    \small
    \begin{tabular}{lrrrrrrrrrrr}
\toprule
{} &  Airplane &  Automobile &   Bird &    Cat &   Deer &    Dog &   Frog &  Horse &   Ship &  Truck &   Avg. \\
\midrule
Blur      &     10.60 &       11.40 &  13.10 &   9.81 &   7.30 &  13.50 &  17.70 &   9.00 &   6.00 &   6.20 &  10.46 \\
Flip L/R  &      2.90 &        1.00 &   4.10 &   6.71 &   2.20 &   6.80 &   1.30 &   2.40 &   0.90 &   2.40 &   3.07 \\
Flip U/D  &     14.90 &       74.60 &  37.80 &  33.13 &  59.10 &  53.90 &  29.30 &  92.40 &  72.20 &  43.30 &  51.06 \\
Grayscale &      4.70 &        5.40 &  28.10 &   7.91 &  18.10 &  26.00 &  14.30 &   6.70 &   4.80 &   5.30 &  12.13 \\
Invert    &     16.50 &       29.40 &  29.50 &  33.13 &  41.40 &  70.30 &  41.80 &  38.30 &  27.30 &  35.70 &  36.33 \\
Rotation  &     25.49 &       37.09 &  35.43 &  17.70 &  69.00 &  46.10 &  20.63 &  60.44 &  42.44 &  50.01 &  40.43 \\
Shear     &     11.22 &        4.99 &  26.69 &  35.79 &  45.45 &  51.97 &  15.63 &  40.24 &  19.78 &  55.24 &  30.70 \\
Avg.      &     12.33 &       23.41 &  24.96 &  20.60 &  34.65 &  38.37 &  20.10 &  35.64 &  24.77 &  28.31 &  26.31 \\
\bottomrule
\end{tabular}

    \caption{CIFAR-10 dataset: Effects of MRs by the true class of the image. Each cell value shows the percentage of images in the class, which are wrongly classified after applying the MR. Every class contains 1000 images. Rotation and Shear are parameterized by $30$ degrees. The values are determined using the baseline method, i.e. exhaustive search over all MRs and all images. The rightmost column (Avg.) corresponds to the baseline results in the image classification case study (see Figure~\ref{fig:cs1a_main}).}
    \label{tab:effects_per_action}
\end{table}

\subsection{Fault-Revealing Metamorphic Relations}

\subsubsection{Case Study 1: Image Classification}

In our first case study, which is an image classification ML model, the goal of the bandit is to select a MR which leads to a different classification of the transformed image compared to the original image of the source test case.
While we looked at the true classes of the source images in the previous initial experiment, we only consider the change of the predicted class from source to a follow-up test case in this experiment.
This approach focuses on the consistency of outputs for different variants of the same input image, i.e., the follow-up test cases.
The actual true label is not as relevant in this case, as it is unlikely to find one MR which transforms the image consistently in a way that the correct label is predicted.
Additionally, the accuracy, i.e., the correctness of the outputs is usually already tested during or after the training of the model.
Focusing on the difference between outputs for source and follow-up test cases furthermore allows to extend the set of source test cases from unlabelled datasets, making it easier to enlarge the system's test suite.

\begin{figure}[t]
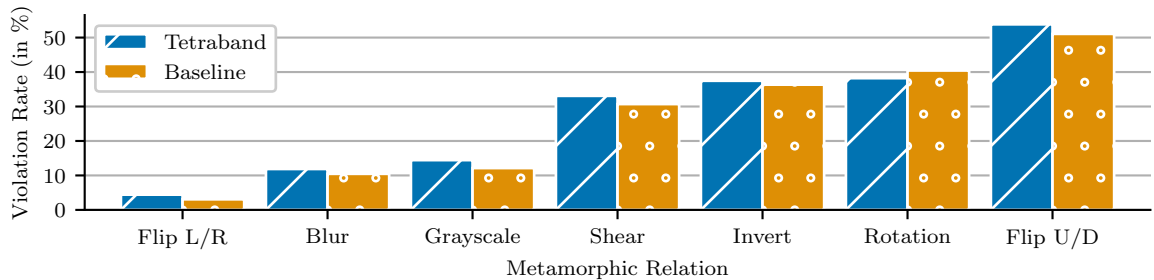
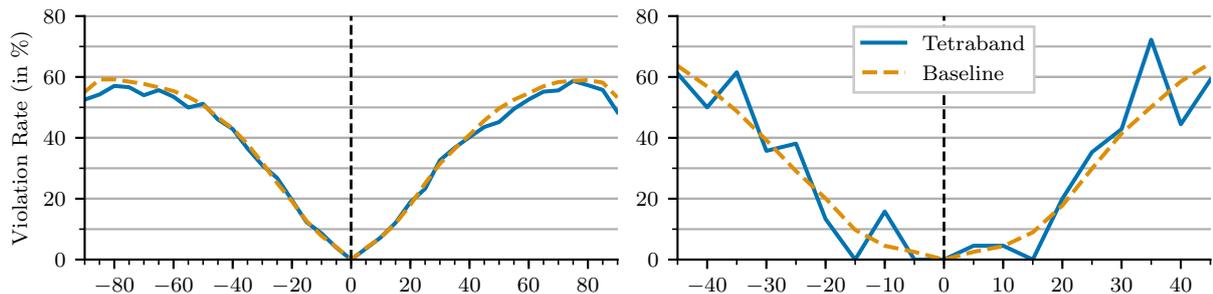

    \centering
    \begin{subfigure}[t]{\textwidth}
      \input{figures/classification-cifar10-hierarchical-main-bandit-actions.pgf}
      \caption{Violation rate of MRs selected by \method and baseline results for exhaustive search}
      \label{fig:cs1a_main}
    \end{subfigure}\\
    \begin{subfigure}[t]{0.5\textwidth}
      \input{figures/classification-cifar10-hierarchical-rotation-bandit-actions-wide.pgf}
      \caption{Configuration of Rotation transformation (in degrees)}
      \label{fig:cs1a_rotation}
    \end{subfigure}\hfill
    \begin{subfigure}[t]{0.5\textwidth}
      \input{figures/classification-cifar10-hierarchical-shear-bandit-actions-wide.pgf}
      \caption{Configuration of Shear transformation (in degrees)}
      \label{fig:cs1a_shear}
    \end{subfigure}
    \caption{Fault-Revealing MRs for image classification: Violation rate and
          configuration of parameterized MR transformations.
          \method approximates the true error distribution and select fault-revealing MRs and their parameters.}
    \label{fig:basic_actions}
\end{figure}

For the seven main transformations, \figurename~\ref{fig:cs1a_main} shows the distribution of the violation rate for each follow-up test case generated by a selected MR. 
We compare this violation rate of MRs selected by \method to the true violation rate for the set of source test cases, which corresponds to the rightmost column (Avg.) in Table~\ref{tab:effects_per_action}. 
These ground-truth results form our baseline and are determined by applying all available MRs to all source test cases in an exhaustive search.
While this exhaustive search covers all possible follow-up test cases, it is time- and resource-intense compared to adaptive metamorphic testing and not ideally suited for repeated testing.
We have therefore considered the random selection of a MR, such as it is common in traditional metamorphic testing, as the comparison method for computational cost and resource consumption. 
Conclusively, to discuss the quality of the selected MRs with \method, we use exhaustive search as the baseline.
The evaluation of computational cost, including a comparison to the common random selection, is discussed separately in Section~\ref{sec:compcost}.

The \textit{violation rate} estimates how often the image classifier changes its prediction after the MR was applied to the source image.
This can be different from the \textit{failure rate} in cases where the prediction on the original image was wrong, but the transformation made the model predict the correct output.
However, as we do not expect the test data to be labeled, we do not require and consider this information for our experiments and instead focus on the robustness and consistency of the model for the predictions. Therefore, we mostly report the violation rate.
Ideally, a perfectly-trained image classification model should not show any change and the violation rates should be zero independently of the selected MR.

Generally speaking, the violation rate distribution shows a different impact for different MRs.
The least fault-revealing MR is the one that mirrors images over the middle vertical axis, i.e., flip left and right.
This transformation is most likely to be found in the training datasets for the image classifier. 
Often the image is either symmetric by itself, like a human face or body, or included together with another image of the same object but taken from a different angle, showing a symmetric profile.
Other MRs are more effective to reveal faults in the image classifier, which is related to the disturbance impact they have on the original image. 
They often represent transformations that could be expected in real-world applications and should be covered by a robust image classification system.
As seen from the exhaustive search baseline experiment, the most fault-revealing MR identified by \method showed to be flipping the image upside-down, i.e., mirroring on the middle of the horizontal axis.
This is within expectations, as it results in an image, which is unlikely to be represented in the distribution of training images, e.g., an image of a car is likely to be shown with its wheels on the ground.
Nevertheless, high violation rates are also observed for less invasive MRs, such as rotating the image or inverting it, and these MRs are either likely to be encountered in practical applications or preserve many of the distinctive features.
Having this statistic not only allows us to identify the weaknesses of the classifier (model testing), but it can also be used as a basis to configure image augmentation techniques to train a new version of the image classification model (model training).
With image augmentation, the training set of images is extended by including modified versions of the original image, through small perturbations or affinity scaling, while preserving the original label.
By knowing the MRs that fail the old classifier, the necessary transformations to include in future image augmentation are known and can help to improve the performance of new models.
However, not all MRs are necessarily suitable image augmentations at training time, as they might produce images that are not within the distribution of inputs, i.e. images, for which the model is trained.

The performance of \method closely approximates the violation distribution for the baseline results and exceeds them for all MRs, except Rotation where the violation rate is close to the baseline violation rate. 
The reason for the lower violation rate is related to the necessary exploration to select an appropriate parameter to rotate the source image. 
Due to the large number of parameters from $-90$ to $90$ the bandit exploration had to take ineffective actions to learn, which leads to an initially lower violation rate. 

However, we also observe that longer runtime and more iterations over the dataset further increase the averaged violation rate as the bandit algorithm can focus more on exploitation than exploration.
This can be explained as initial iterations do not have knowledge about the MRs effectiveness and the selection is more random, which includes selecting ineffective MRs.
While these actions are valuable for exploration, they do not contribute to the violation rate.
Later, when the effect of MRs has been sufficiently explored, the focus changes on also exploiting the MRs and selecting fault-revealing MRs.
These actions then contribute to increasing the violation rate. Accordingly, if the system under test does not rapidly change, more iterations will increase the amount of exploitation with high violation rates and the impact of the initial exploration on the violation rate decreases.

The violation distribution for the different parameters of the MRs Rotation and Shear are shown in \figurename~\ref{fig:cs1a_rotation} and \figurename~\ref{fig:cs1a_shear}.
The bandit effectively picks the appropriate degree of rotation to closely resemble the true error distribution. 
The only exceptions are the largest degrees of rotation, where the selection does not completely approximate the true distribution.
However, due to our reward structure, the agent is encouraged to focus on minimal rotations in the images that lead to some misclassification. 
Especially in the range between $-45$ and $45$ degrees the parameter selection is appropriate, which indicates the successful convergence towards the most revealing parameter for image rotation.

In the Shear transformation, \method was less effective to smoothly approximate the true error distribution but broadly follows its shape. 
Here, we observe that due to the lower general violation rate of the MR, there are fewer chances for successful exploration of the parameter space than for the Rotation MR.
Accordingly, the approximation of the parameter distribution for Shear is not as close as for Rotation, but still follows the general distribution.

\subsubsection{Case Study 2: Object Detection}

\begin{figure}[t]
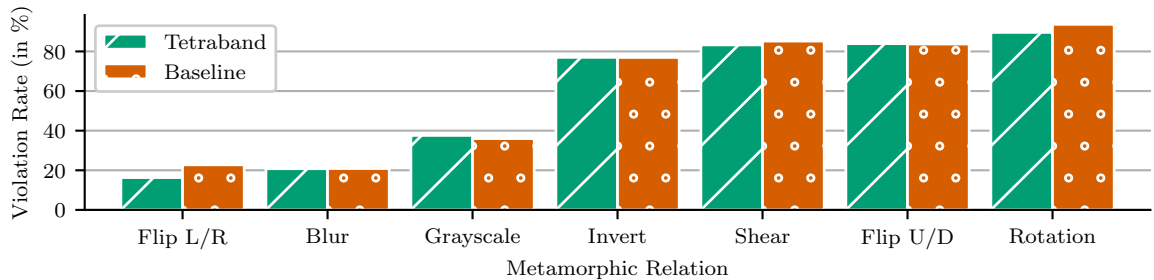
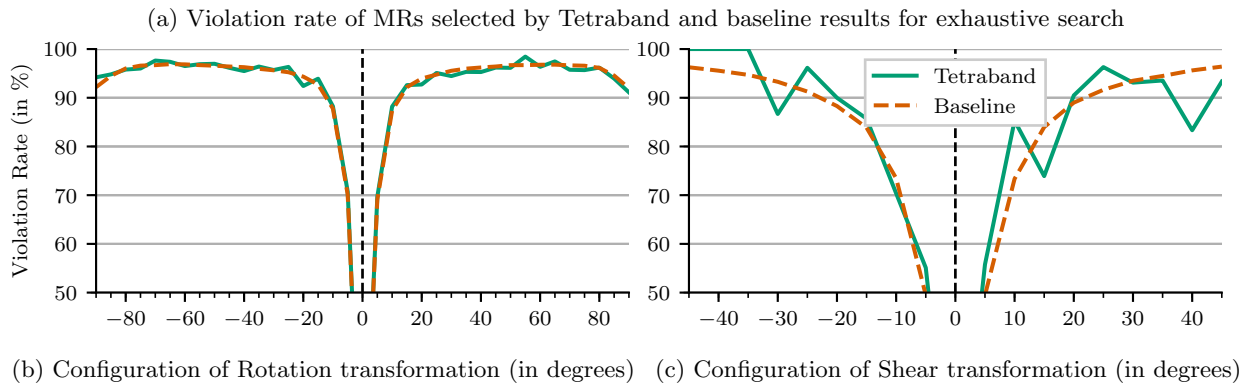

    \centering
    \begin{subfigure}[t]{\textwidth}
      \input{figures/detection-coco-hierarchical-main-bandit-actions.pgf}
      \caption{Violation rate of MRs selected by \method and baseline results for
    exhaustive search}
      \label{fig:cs1b_main}
    \end{subfigure}\\
    \begin{subfigure}[t]{0.5\textwidth}
      \input{figures/detection-coco-hierarchical-rotation-bandit-actions-wide.pgf}
      \caption{Configuration of Rotation transformation (in degrees)}
      \label{fig:cs1b_rotation}
    \end{subfigure}\hfill
    \begin{subfigure}[t]{0.5\textwidth}
      \input{figures/detection-coco-hierarchical-shear-bandit-actions-wide.pgf}
      \caption{Configuration of Shear transformation (in degrees)}
      \label{fig:cs1b_shear}
    \end{subfigure}
    \caption{Fault-Revealing MRs for object detection: Violation rate and
          configuration of parameterized MR transformations.
          \method approximates the true error distribution and select fault-revealing MRs and their parameters.}
    \label{fig:basic_actions_detection}
\end{figure}

The second case study application is the object detection neural network.
Similar to the presentation of the first case study, \figurename~\ref{fig:basic_actions_detection} shows the results of the object detection case study.
As a first main difference, the results show a higher violation rate for object detection, with over 70\% violation rate for four of the seven MRs.
The ranking of MRs is similar but applying the Shear MR with parameter
selection is more efficient here than in the image classification case study,
where the Invert MR showed a higher violation rate, and Rotation is the most effective MR.
The MRs Flip L/R, Blur and Grayscale have the lowest violation rate, i.e., the model is most robust to these changes.
However, while the total violation rate is higher, the case study itself is more difficult as the dataset on which we apply \method is smaller and only consists of 5000 images, which means the overall process takes half of the iterations of the image classification case study. 
This difference explains the approximation difference for some of the main MRs in comparison to the baseline violation rates.

For the two MRs Rotation and Shear, the approximation quality for the additional parameter is similar to the image classification results.
For Rotation, the distribution is close to the true distribution of the exhaustive search baseline results, related to the high overall violation rate for this MR and its corresponding higher selection and thereby better exploration opportunities.
The Shear parameters match the true distribution less closely in this case study, which is also related to the higher general effectiveness of this MR for the object detection dataset.

In general, the results of the object detection case study confirm the results of the image classification case study while at the same time respecting the higher difficulty of fewer iterations, due to which the violation rate of the main MRs is close to the true violation rate, but does not exceed it after the given number of iterations.

\subsubsection{Computational Cost and Random Selection}
\label{sec:compcost}

From the previous experiments, we have evaluated the effectiveness of \method for selecting fault-revealing MRs in image classification and object detection systems.
However, we did not consider the computational cost of introducing machine learning in the MT process or compared \method to the commonly used approach of randomly selecting MRs and their parameters.
In this experiment, to answer RQ3, we analyze common characteristics for both case studies, which consider the computational cost of introduced contextual bandits and the additional learning step in the MR process. 
We further briefly discuss another comparison method that more closely resembles the state-of-practice in MT, which is to randomly select MRs and their parameters.
A summary of the results is given in Table~\ref{tab:compcost}.

\begin{table}[t]
    \centering
    \begin{tabular}{lrrrrr}
        \toprule
                        & \multicolumn{2}{c}{Runtime (in s)} & \multicolumn{3}{c}{Accuracy (in \%)} \\
                        \cmidrule(r){2-3} \cmidrule(l){4-6}
            Case Study  & \method & Random & Unmodified & \method & Random\\
        \midrule
            Image Classification & 0.1 & 0.06 & 96.6 & 54.0 & 72.9 \\
            Object Detection & 0.63 & 0.56 & 54.3 & 23.1 & 36.3 \\
        \bottomrule     
    \end{tabular}
    \caption{Computational cost and accuracy of \method and random selection for selecting fault-revealing MRs. The average runtime per image shows that the overhead introduced by the ML model is relatively small, especially for object detection with a more costly test execution. \method selects MRs more effectively and the accuracy is smaller than with random selection.}
    \label{tab:compcost}
\end{table}

Running \method is computationally cheaper and more sample-efficient than the exhaustive search that we consider as a baseline.
While the exhaustive search considers all MRs with all different parameters for Rotation and Shear, in total 59 different transformations, \method selects one MR and one parameter per iteration.
In addition to the application of the MR and the execution of the SUT, \method has a small computational overhead for selecting the MR and its parameter and learning from the received feedback. 
The average duration per iteration in the image classification case study is 0.1s using \method and 0.06s when using a random MR and not learning from feedback. 
While this overhead increases the execution time, it is still faster and more efficient than running all possible transformations, as we will see below.
For object detection, the average duration is 0.63s with learning and 0.56s without, here the main computation lies in the object detection neural network.
The overhead for training the contextual bandit could be further reduced by moving the learning step outside the main processing loop; however, we argue that in a practical application the overhead is negligible due to the lower number of executions and the availability of highly optimized contextual bandit implementations.

We also considered a random selection of MRs and their parameters. 
At each iteration, a random MR is sampled uniformly instead of using the bandit selection.
This selection is less efficient than \method. 
Using \method, the accuracy of the image classifier is reduced from 96.6\% for the unmodified images to 54.0\% for the images modified by the selected MRs. 
With random selection, the accuracy for the modified images remains at 72.9\%, which is a substantial reduction, but not as high as \method. 
In object detection, the precision is reduced from 54.3\% to 23.1\% with \method and to 36.3\% with random selection.
Due to the lower general effectiveness of random selection, we do not further discuss its results in detail.

\subsection{Robustness Boundaries}

\begin{figure}[t]
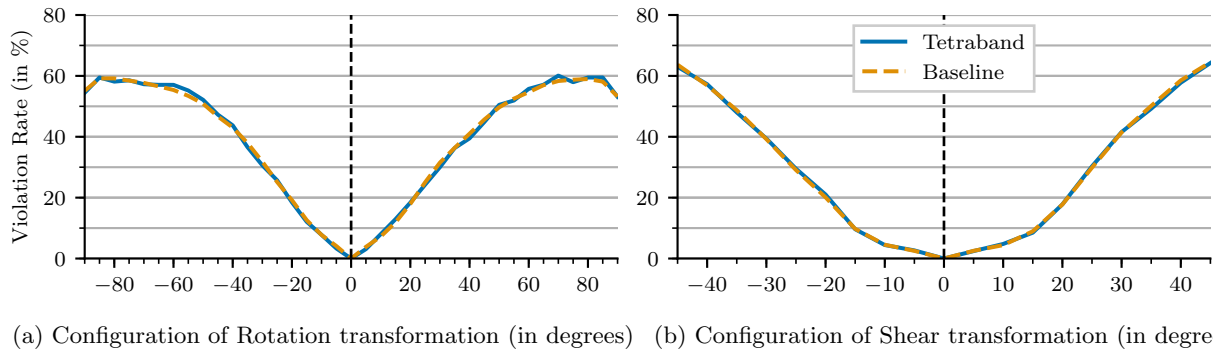

  \centering
  \begin{subfigure}[t]{0.5\textwidth}
    \input{figures/classification-cifar10-rotation-rotation-bandit-actions-wide.pgf}
    \caption{Configuration of Rotation transformation (in degrees)}
    \label{fig:rotation_classification}
  \end{subfigure}\hfill
  \begin{subfigure}[t]{0.5\textwidth}
    \input{figures/classification-cifar10-shear-shear-bandit-actions-wide.pgf}
    \caption{Configuration of Shear transformation (in degrees)}
    \label{fig:shear_classification}
  \end{subfigure}
  \caption{Image classification: Average violation rate per degree step. The approximated violation rate closely approximates the true violation distribution of the ground truth baseline, i.e. exhaustive search.}
  \label{fig:robustness_classification}
\end{figure}

\begin{figure}[t]
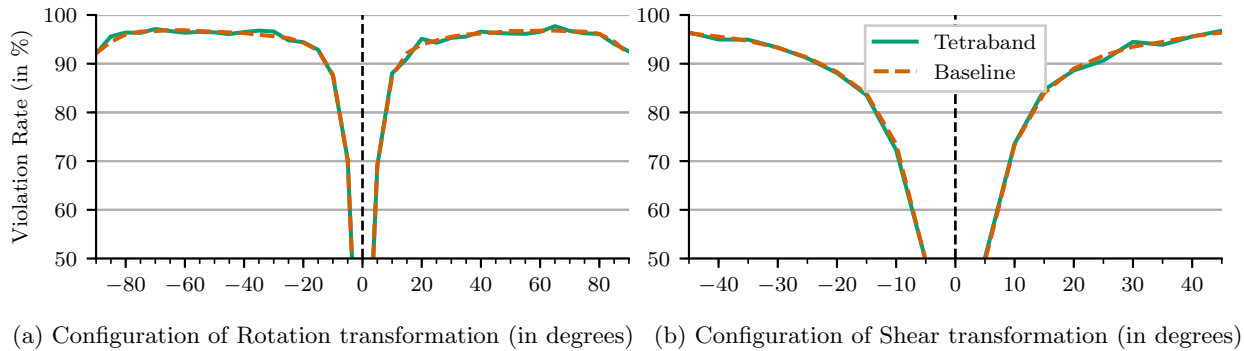

  \centering
  \begin{subfigure}[t]{0.5\textwidth}
    \input{figures/detection-coco-rotation-rotation-bandit-actions-wide.pgf}
    \caption{Configuration of Rotation transformation (in degrees)}
    \label{fig:rotation_detection}
  \end{subfigure}\hfill
  \begin{subfigure}[t]{0.5\textwidth}
    \input{figures/detection-coco-shear-shear-bandit-actions-wide.pgf}
    \caption{Configuration of Shear transformation (in degrees)}
    \label{fig:shear_detection}
  \end{subfigure}
  \caption{Object detection: Average violation rate per degree step. The approximated violation rate closely approximates the true violation distribution of the ground truth baseline, i.e. exhaustive search.}
  \label{fig:robustness_detection}
\end{figure}

As a second experiment and application of \method to learning MR selection, we aim to find the robustness boundaries of the case study systems against different degrees of image rotations and shearing. 
We show the results for both case study applications in \figurename~\ref{fig:robustness_classification} for image classification and \figurename~\ref{fig:robustness_detection} for object detection. 
The experimental results mostly confirm the inherent hypothesis, also following the previous results from the first experiments that larger modifications of the source test case lead to a higher violation rate of the follow-up test case. 
This is true for both case study applications, but the extent to which the effect applies varies with the object detection system being much more susceptible to images rotated even only by small degrees.
At the same time, we confirm that a larger number of iterations allows better exploration and approximation of the true error distribution. 
Where the parameter distribution in the previous experiments showed divergence, mostly for larger parameter values, the focus on the specific MRs in this experiment allows sufficient exploration and good approximation.

The results for image classification clearly show the effect that larger rotations of the original image are more likely to cause a different classification (see \figurename~\ref{fig:rotation_classification}). 
Here, the bandit does not only learn to select the largest rotation but does accurately approximate the distribution of true faults as given by the exhaustive search baseline. 
Our results show that for more than 10\% of the source test cases a rotation of at least 10 degrees leads to a different image classification. 
When considering real-world scenarios for the application of image classification systems, a rotation of 10 degrees can likely occur due to tilt or shifts in either the camera or due to external influences on the actual object.

For object detection, the interpretation of the results needs to consider two aspects, which lead to the conclusion that the results can not be directly compared to the image classification case study. 
First, the original SUT already has lower performance for the original data set than the image classification SUT. 
When considering the system to be more imprecise for unmodified data, then it is also likely to be more fragile for modified data. 
Second, the evaluation metric used, mean average precision, is more fragile than the metric used for image classification. 
The results show the fragility of the object detection system, as well as the capability to learn to approximate this error distribution over transforming each of the 5000 source test case images only once.

\subsection{Discussion} \label{sec:discussion}

For both case studies in our experiments, our results showed the effectiveness of contextual bandits, as part of \method, to adapt to a prior unknown error distribution in two different case study applications, based on two different neural network architectures and tasks.

From the results, we draw two major conclusions.
First, we see a confirmation for the applicability of \method for selecting metamorphic relations using contextual bandits (i.e., Adaptive Metamorphic Testing), as is shown by the close approximation of the true error distribution with limited iterations.
Second, our tests reveal robustness weaknesses in the two systems-under-test.
Weaknesses in neural networks have been addressed before and are an active research area \cite{Goodfellow2009,Carlini2017,Biggio2018}.
The research under the area of \emph{adversarial examples} focuses on finding input perturbations that lead to misbehavior of the model with only minimal or hard-to-detect changes in the input.
This approach is different from the setting of our experiment.
We select distinct and known image transformations to modify the image without the goal to hide the transformation, which often is the intent of an adversarial example.

\section{Conclusion} \label{sec:conclusion}

This paper introduces Adaptive Metamorphic Testing (AMT), a method to control metamorphic testing using contextual bandits.
AMT receives a feature vector representing the source test case and selects a MR to generate a follow-up test case.
From the result of evaluating the follow-up test case, whether it reveals a fault, the bandit learns which MRs can be used to exploit weaknesses in the system.
At the same time, using state-of-the-art algorithms for contextual bandits, AMT explores non-optimal actions to identify previously unknown weaknesses and to adapt to changing behavior in repeated testing of the same system.

We have evaluated the applicability of AMT using our implementation \method on two image analysis case studies in two distinct tasks.
For both case studies, our results showed that \method approximates the true distribution of faults in the SUT with fewer iterations and executions of the SUT than an exhaustive search and more efficiently than random sampling of MRs and their parameters, which is the common best practice.
\method learns to select fault-revealing MRs in relation to the source test case while ignoring non-relevant MRs.
Furthermore, in the second experiment, \method proved to be effective for the identification of robustness boundaries, that is, exploring the parameterization of individual MRs, which have different impacts on the SUT.
Our experiment explored how different degrees of rotation and shear affected the classification result of the transformed image.

In conclusion, AMT is effective for selecting MRs and is more time-efficient than exhaustive testing and more effective than the standard approach of pure random sampling.
We see the method to be useful in repeated testing scenarios, such as continuous integration, where regression of the SUT can be tested from an initial knowledge about previous fault characteristics.
For future work, we plan to investigate the combination of multiple MRs to create follow-up test cases instead of selecting only one MR per iteration.

\section*{Acknowledgements}
This work is supported by the Research Council of Norway (RCN) through the research-based innovation center Certus, under the SFI program.
The experiments were performed on the Abel Cluster, owned by the University of Oslo and Uninett/Sigma2, and operated by the Department for Research Computing at USIT, the University of Oslo IT-department.

\section*{Declarations of Interest}
None of the authors declares a conflict of interest.

\bibliography{library}

\begin{thebibliography}{10}
\expandafter\ifx\csname url\endcsname\relax
  \def\url#1{\texttt{#1}}\fi
\expandafter\ifx\csname urlprefix\endcsname\relax\def\urlprefix{URL }\fi
\expandafter\ifx\csname href\endcsname\relax
  \def\href#1#2{#2} \def\path#1{#1}\fi

\bibitem{Chen1998}
T.~Chen, S.~Cheung, S.~Yiu, Metamorphic {{Testing}}: {{A New Approach}} for
  {{Generating Next Test Cases}}, Technical {{Report}} HKUST-CS98-01,
  {Department of Computer Science, Hong Kong University of Science and
  Technology}, {Hong Kong} (1998).

\bibitem{Chen2018}
T.~Y. Chen, F.-C. Kuo, H.~Liu, P.-L. Poon, D.~Towey, T.~H. Tse, Z.~Q. Zhou,
  Metamorphic {{Testing}}: {{A Review}} of {{Challenges}} and
  {{Opportunities}}, ACM Computing Surveys 51~(1) (2018).
\newblock \href {https://doi.org/10.1145/3143561} {\path{doi:10.1145/3143561}}.

\bibitem{Barr2015}
E.~T. Barr, M.~Harman, P.~McMinn, M.~Shahbaz, S.~Yoo, The {{Oracle Problem}} in
  {{Software Testing}}: {{A Survey}}, IEEE Transactions on Software Engineering
  41~(5) (2015) 507--525.
\newblock \href {https://doi.org/10.1109/TSE.2014.2372785}
  {\path{doi:10.1109/TSE.2014.2372785}}.

\bibitem{Murphy2008}
C.~Murphy, G.~Kaiser, L.~Hu, L.~Wu, Properties of {{Machine Learning
  Applications}} for {{Use}} in {{Metamorphic Testing}}, Proceedings of the
  20th International Conference on Software Engineering and Knowledge
  Engineering (SEKE) (2008) 867--872\href {https://doi.org/10.7916/D8XK8PFD}
  {\path{doi:10.7916/D8XK8PFD}}.

\bibitem{Ding2017}
J.~Ding, X.-H. Hu, V.~Gudivada, A {{Machine Learning Based Framework}} for
  {{Verification}} and {{Validation}} of {{Massive Scale Image Data}}, IEEE
  Transactions on Big Data 26~(3) (2017) 1--1.
\newblock \href {https://doi.org/10.1109/TBDATA.2017.2680460}
  {\path{doi:10.1109/TBDATA.2017.2680460}}.

\bibitem{Dwarakanath2018}
A.~Dwarakanath, M.~Ahuja, S.~Sikand, R.~M. Rao, R.~P. J.~C. Bose, N.~Dubash,
  S.~Podder, Identifying implementation bugs in machine learning based image
  classifiers using metamorphic testing, in: Proceedings of the 27th {{ACM
  SIGSOFT International Symposium}} on {{Software Testing}} and {{Analysis}}
  ({{ISSTA}}), 2018, pp. 118--128.
\newblock \href {https://doi.org/10.1145/3213846.3213858}
  {\path{doi:10.1145/3213846.3213858}}.

\bibitem{Zhou2019}
Z.~Q. Zhou, L.~Sun, Metamorphic {{Testing}} of {{Driverless Cars}},
  Communications of the ACM 62~(3) (2019) 61--67.
\newblock \href {https://doi.org/10.1145/3241979} {\path{doi:10.1145/3241979}}.

\bibitem{Zhou2016b}
Z.~Q. Zhou, S.~Xiang, T.~Y. Chen, Metamorphic {{Testing}} for {{Software
  Quality Assessment}}: {{A Study}} of {{Search Engines}}, IEEE Transactions on
  Software Engineering 42~(3) (2016) 264--284.
\newblock \href {https://doi.org/10.1109/TSE.2015.2478001}
  {\path{doi:10.1109/TSE.2015.2478001}}.

\bibitem{Shahri2019}
M.~P. Shahri, M.~Srinivasan, G.~Reynolds, D.~Bimczok, I.~Kahanda, U.~Kanewala,
  Metamorphic {{Testing}} for {{Quality Assurance}} of {{Protein Function
  Prediction Tools}}, in: 2019 {{IEEE International Conference On Artificial
  Intelligence Testing}} ({{AITest}}), 2019, pp. 140--148.
\newblock \href {https://doi.org/10.1109/AITest.2019.00017}
  {\path{doi:10.1109/AITest.2019.00017}}.

\bibitem{LeCun2015}
Y.~LeCun, Y.~Bengio, G.~Hinton, Deep learning, Nature 521~(7553) (2015)
  436--444.
\newblock \href {https://doi.org/10.1038/nature14539}
  {\path{doi:10.1038/nature14539}}.

\bibitem{Pouyanfar2018}
S.~Pouyanfar, S.~Sadiq, Y.~Yan, H.~Tian, Y.~Tao, M.~P. Reyes, M.-L. Shyu, S.-C.
  Chen, S.~S. Iyengar, A {{Survey}} on {{Deep Learning}}: {{Algorithms}},
  {{Techniques}}, and {{Applications}}, ACM Computing Surveys 51~(5) (2018)
  1--36.
\newblock \href {https://doi.org/10.1145/3234150} {\path{doi:10.1145/3234150}}.

\bibitem{Langford2007}
J.~Langford, T.~Zhang, The {{Epoch}}-{{Greedy Algorithm}} for {{Multi}}-armed
  {{Bandits}} with {{Side Information}}, in: Advances in {{Neural Information
  Processing Systems}} 20 ({{NeurIPS}} 2007), 2007, pp. 817--824.

\bibitem{Zhou2016}
L.~Zhou, A {{Survey}} on {{Contextual Multi}}-armed {{Bandits}}, arXiv preprint
  arXiv:1508.03326 (2016).

\bibitem{Agarwal2014}
A.~Agarwal, D.~Hsu, S.~Kale, J.~Langford, L.~Li, L.~G. Oct, Taming the
  {{Monster}}: {{A Fast}} and {{Simple Algorithm}} for {{Contextual Bandits}},
  in: International {{Conference}} on {{Machine Learning}}, 2014, pp.
  1638--1646.

\bibitem{Lattimore2019}
T.~Lattimore, C.~Szepesvari, Bandit {{Algorithms}}, Vol. Revision:
  8b22b8b6131c37e388d5e3b2eecf0b4ff5d7db92, {https://banditalgs.com/}, 2019.

\bibitem{Sutton2018}
R.~S. Sutton, A.~G. Barto, Reinforcement {{Learning}}: {{An Introduction}}, 2nd
  Edition, {MIT Press}, 2018.

\bibitem{Li2010}
L.~Li, W.~Chu, J.~Langford, R.~E. Schapire, A contextual-bandit approach to
  personalized news article recommendation, in: International {{Conference}} on
  {{World Wide Web}} ({{WWW}}), 2010, pp. 661--670.
\newblock \href {https://doi.org/10.1145/1772690.1772758}
  {\path{doi:10.1145/1772690.1772758}}.

\bibitem{Lu2010}
T.~Lu, D.~Pal, M.~Pal, Contextual {{Multi}}-{{Armed Bandits}}, in: Proceedings
  of the 13th {{International}} {{Conferenceon Artificial Intelligence}} and
  {{Statistics}} ({{AISTATS}}), 2010, pp. 485--492.

\bibitem{Tang2013}
L.~Tang, R.~Rosales, A.~Singh, D.~Agarwal, Automatic ad format selection via
  contextual bandits, in: Proceedings of the 22nd {{ACM}} International
  Conference on {{Conference}} on Information \& Knowledge Management, 2013,
  pp. 1587--1594.
\newblock \href {https://doi.org/10.1145/2505515.2514700}
  {\path{doi:10.1145/2505515.2514700}}.

\bibitem{Baskiotis2006}
N.~Baskiotis, M.~Sebag, M.-C. Gaudel, S.-D. Gouraud, {{EXIST}}:
  {{Exploitation}}/{{Exploration Inference}} for {{Statistical Software
  Testing}}, in: On-Line {{Trading}} of {{Exploration}} and {{Exploitation}},
  {{NeurIPS}} {{Workshop}}, 2006.

\bibitem{Loth2013}
M.~Loth, M.~Sebag, Y.~Hamadi, M.~Schoenauer, Bandit-based {{Search}} for
  {{Constraint Programming}}, in: International {{Conference}} on
  {{Principles}} and {{Practice}} of {{Constraint Programming}}, 2013, pp.
  464--480.

\bibitem{Balafrej2015}
A.~Balafrej, C.~Bessiere, A.~Paparrizou, Multi-armed bandits for adaptive
  constraint propagation, International Joint Conference on Artificial
  Intelligence (2015) 290--296.

\bibitem{Ontanon2017}
S.~Onta{\~n}{\'o}n, Combinatorial {{Multi}}-armed {{Bandits}} for
  {{Real}}-{{Time Strategy Games}}, Journal of Artificial Intelligence Research
  58~(1) (2017) 665--702.
\newblock \href {https://doi.org/10.1613/jair.5398}
  {\path{doi:10.1613/jair.5398}}.

\bibitem{Segura2016}
S.~Segura, G.~Fraser, A.~B. Sanchez, A.~{Ruiz-Cortes}, A {{Survey}} on
  {{Metamorphic Testing}}, IEEE Transactions on Software Engineering 42~(9)
  (2016) 805--824.
\newblock \href {https://doi.org/10.1109/TSE.2016.2532875}
  {\path{doi:10.1109/TSE.2016.2532875}}.

\bibitem{Sun2018}
L.~Sun, Z.~Q. Zhou, Metamorphic {{Testing}} for {{Machine Translations}}:
  {{MT4MT}}, in: 2018 25th {{Australasian Software Engineering Conference}}
  ({{ASWEC}}), 2018, pp. 96--100.
\newblock \href {https://doi.org/10.1109/ASWEC.2018.00021}
  {\path{doi:10.1109/ASWEC.2018.00021}}.

\bibitem{Segura2018}
S.~Segura, J.~Troya, A.~Dur{\'a}n, A.~{Ruiz-Cort{\'e}s}, Performance
  metamorphic testing: {{A Proof}} of concept, Information and Software
  Technology 98 (2018) 1--4.
\newblock \href {https://doi.org/10.1016/j.infsof.2018.01.013}
  {\path{doi:10.1016/j.infsof.2018.01.013}}.

\bibitem{Johnston2019}
O.~Johnston, D.~Jarman, J.~Berry, Z.~Q. Zhou, T.~Y. Chen, Metamorphic
  {{Relations}} for {{Detection}} of {{Performance Anomalies}}, in: 2019
  {{IEEE}}/{{ACM}} 4th {{International Workshop}} on {{Metamorphic Testing}}
  ({{MET}}), 2019, pp. 63--69.
\newblock \href {https://doi.org/10.1109/MET.2019.00017}
  {\path{doi:10.1109/MET.2019.00017}}.

\bibitem{Akgun2018}
{\"O}.~Akg{\"u}n, I.~P. Gent, C.~Jefferson, I.~Miguel, P.~Nightingale,
  Metamorphic {{Testing}} of {{Constraint Solvers}}, in: J.~Hooker (Ed.),
  Principles and {{Practice}} of {{Constraint Programming}}, Vol. 11008 of
  {{LNCS}}, 2018, pp. 727--736.
\newblock \href {https://doi.org/10.1007/978-3-319-98334-9\_46}
  {\path{doi:10.1007/978-3-319-98334-9\_46}}.

\bibitem{Gotlieb2003}
A.~Gotlieb, B.~Botella, Automated metamorphic testing, Proceedings 27th Annual
  International Computer Software and Applications Conference (2003)
  34--40\href {https://doi.org/10.1109/CMPSAC.2003.1245319}
  {\path{doi:10.1109/CMPSAC.2003.1245319}}.

\bibitem{Mayer2006}
J.~Mayer, R.~Guderlei, An {{Empirical Study}} on the {{Selection}} of {{Good
  Metamorphic Relations}}, in: 30th {{Annual International Computer Software}}
  and {{Applications Conference}}, 2006, pp. 475--484.
\newblock \href {https://doi.org/10.1109/COMPSAC.2006.24}
  {\path{doi:10.1109/COMPSAC.2006.24}}.

\bibitem{Kanewala2013}
U.~Kanewala, J.~M. Bieman, Using machine learning techniques to detect
  metamorphic relations for programs without test oracles, 2013 IEEE 24th
  International Symposium on Software Reliability Engineering (ISSRE) (2013).
\newblock \href {https://doi.org/10.1109/ISSRE.2013.6698899}
  {\path{doi:10.1109/ISSRE.2013.6698899}}.

\bibitem{Kanewala2016}
U.~Kanewala, J.~M. Bieman, A.~{Ben-Hur}, Predicting metamorphic relations for
  testing scientific software: A machine learning approach using graph kernels,
  Software Testing, Verification and Reliability 26~(3) (2016) 245--269.
\newblock \href {https://doi.org/10.1002/stvr.1594}
  {\path{doi:10.1002/stvr.1594}}.

\bibitem{Pei2017}
K.~Pei, Y.~Cao, J.~Yang, S.~Jana, {{DeepXplore}}: {{Automated Whitebox
  Testing}} of {{Deep Learning Systems}}, in: Proceedings of the 26th
  {{Symposium}} on {{Operating Systems Principles}}, {ACM Press}, 2017, pp.
  1--18.
\newblock \href {https://doi.org/10.1145/3132747.3132785}
  {\path{doi:10.1145/3132747.3132785}}.

\bibitem{Ma2018}
L.~Ma, Y.~Liu, J.~Zhao, Y.~Wang, F.~{Juefei-Xu}, F.~Zhang, J.~Sun, M.~Xue,
  B.~Li, C.~Chen, T.~Su, L.~Li, {{DeepGauge}}: Multi-granularity testing
  criteria for deep learning systems, in: Proceedings of the 33rd
  {{ACM}}/{{IEEE International Conference}} on {{Automated Software
  Engineering}} - {{ASE}} 2018, 2018, pp. 120--131.
\newblock \href {https://doi.org/10.1145/3238147.3238202}
  {\path{doi:10.1145/3238147.3238202}}.

\bibitem{Xie2011}
X.~Xie, J.~W. Ho, C.~Murphy, G.~Kaiser, B.~Xu, T.~Y. Chen, Testing and
  validating machine learning classifiers by metamorphic testing, Journal of
  Systems and Software 84~(4) (2011) 544--558.
\newblock \href {https://doi.org/10.1016/j.jss.2010.11.920}
  {\path{doi:10.1016/j.jss.2010.11.920}}.

\bibitem{Chan2010}
W.~K. Chan, J.~C.~F. Ho, T.~H. Tse, Finding failures from passed test cases:
  Improving the pattern classification approach to the testing of mesh
  simplification programs, Software Testing, Verification and Reliability
  20~(2) (2010) 89--120.
\newblock \href {https://doi.org/10.1002/stvr.408}
  {\path{doi:10.1002/stvr.408}}.

\bibitem{Yang2019}
S.~Yang, D.~Towey, Z.~Q. Zhou, Metamorphic {{Exploration}} of an {{Unsupervised
  Clustering Program}}, in: Proceedings of the 4th {{International Workshop}}
  on {{Metamorphic Testing}}, 2019, pp. 48--54.
\newblock \href {https://doi.org/10.1109/MET.2019.00015}
  {\path{doi:10.1109/MET.2019.00015}}.

\bibitem{Mekala2019}
R.~R. Mekala, G.~E. Magnusson, A.~Porter, M.~Lindvall, M.~Diep, Metamorphic
  {{Detection}} of {{Adversarial Examples}} in {{Deep Learning Models}} with
  {{Affine Transformations}}, in: Proceedings of the 4th {{International
  Workshop}} on {{Metamorphic Testing}}, 2019, pp. 55--62.
\newblock \href {https://doi.org/10.1109/MET.2019.00016}
  {\path{doi:10.1109/MET.2019.00016}}.

\bibitem{Saha2019}
P.~Saha, U.~Kanewala, Fault {{Detection Effectiveness}} of {{Metamorphic
  Relations Developed}} for {{Testing Supervised Classifiers}}, in: 2019 {{IEEE
  International Conference On Artificial Intelligence Testing}} ({{AITest}}),
  2019, pp. 157--164.
\newblock \href {https://doi.org/10.1109/AITest.2019.00019}
  {\path{doi:10.1109/AITest.2019.00019}}.

\bibitem{Cai2007}
K.-Y. Cai, B.~Gu, H.~Hu, Y.-C. Li, Adaptive software testing with fixed-memory
  feedback, Journal of Systems and Software 80~(8) (2007) 1328--1348.
\newblock \href {https://doi.org/10.1016/j.jss.2006.11.008}
  {\path{doi:10.1016/j.jss.2006.11.008}}.

\bibitem{Zhou2018}
Z.~Q. Zhou, A.~Sinaga, W.~Susilo, L.~Zhao, K.-Y. Cai, A cost-effective software
  testing strategy employing online feedback information, Information Sciences
  422 (2018) 318--335.
\newblock \href {https://doi.org/10.1016/j.ins.2017.08.088}
  {\path{doi:10.1016/j.ins.2017.08.088}}.

\bibitem{Mayer2006a}
J.~Mayer, R.~Guderlei, On {{Random Testing}} of {{Image Processing
  Applications}}, in: 2006 {{Sixth International Conference}} on {{Quality
  Software}} ({{QSIC}}'06), 2006, pp. 85--92.
\newblock \href {https://doi.org/10.1109/QSIC.2006.45}
  {\path{doi:10.1109/QSIC.2006.45}}.

\bibitem{Guderlei2007}
R.~Guderlei, J.~Mayer, Towards automatic testing of imaging software by means
  of random and metamorphic testing, International Journal of Software
  Engineering and Knowledge Engineering 17~(06) (2007) 757--781.
\newblock \href {https://doi.org/10.1142/S0218194007003471}
  {\path{doi:10.1142/S0218194007003471}}.

\bibitem{Xu2018}
L.~Xu, D.~Towey, A.~P. French, S.~Benford, Z.~Q. Zhou, T.~Y. Chen, Enhancing
  supervised classifications with metamorphic relations, in: Proceedings of the
  3rd {{International Workshop}} on {{Metamorphic Testing}} - {{MET}} '18,
  2018, pp. 46--53.
\newblock \href {https://doi.org/10.1145/3193977.3193978}
  {\path{doi:10.1145/3193977.3193978}}.

\bibitem{Russakovsky2015}
O.~Russakovsky, J.~Deng, H.~Su, J.~Krause, S.~Satheesh, S.~Ma, Z.~Huang,
  A.~Karpathy, A.~Khosla, M.~Bernstein, A.~C. Berg, L.~{Fei-Fei}, {{ImageNet
  Large Scale Visual Recognition Challenge}}, International Journal of Computer
  Vision 115~(3) (2015) 211--252.
\newblock \href {https://doi.org/10.1007/s11263-015-0816-y}
  {\path{doi:10.1007/s11263-015-0816-y}}.

\bibitem{Liu2016a}
W.~Liu, D.~Anguelov, D.~Erhan, C.~Szegedy, S.~Reed, C.-y. Fu, A.~C. Berg,
  {{SSD}}: {{Single Shot MultiBox Detector}}, in: European {{Conference}} on
  {{Computer Vision}}, Vol. 9905 of {{LNCS}}, 2016, pp. 21--37.
\newblock \href {https://doi.org/10.1007/978-3-319-46448-0\_2}
  {\path{doi:10.1007/978-3-319-46448-0\_2}}.

\bibitem{He2016}
K.~He, X.~Zhang, S.~Ren, J.~Sun, Identity mappings in deep residual networks,
  in: European {{Conference}} on {{Computer Vision}}, Vol. 9908 of {{LNCS}},
  2016, pp. 630--645.
\newblock \href {https://doi.org/10.1007/978-3-319-46493-0\_38}
  {\path{doi:10.1007/978-3-319-46493-0\_38}}.

\bibitem{Iandola2016}
F.~N. Iandola, S.~Han, M.~W. Moskewicz, K.~Ashraf, W.~J. Dally, K.~Keutzer,
  {{SqueezeNet}}: {{AlexNet}}-level accuracy with 50x fewer parameters and
  {$<$}0.{{5MB}} model size, arXiv preprint arXiv:1602.07360 (2016).

\bibitem{Krizhevsky2014a}
A.~Krizhevsky, V.~Nair, G.~Hinton, The {{CIFAR}}-10 dataset, online:
  http://www. cs. toronto. edu/kriz/cifar. html (2014).

\bibitem{Huang2017b}
J.~Huang, V.~Rathod, C.~Sun, M.~Zhu, A.~Korattikara, A.~Fathi, I.~Fischer,
  Z.~Wojna, Y.~Song, S.~Guadarrama, K.~Murphy, Speed/{{Accuracy
  Trade}}-{{Offs}} for {{Modern Convolutional Object Detectors}}, in: {{IEEE
  Conference}} on {{Computer Vision}} and {{Pattern Recognition}} ({{CVPR}}),
  2017, pp. 3296--3297.
\newblock \href {https://doi.org/10.1109/CVPR.2017.351}
  {\path{doi:10.1109/CVPR.2017.351}}.

\bibitem{Lin2017b}
T.-Y. Lin, P.~Dollar, R.~Girshick, K.~He, B.~Hariharan, S.~Belongie, Feature
  {{Pyramid Networks}} for {{Object Detection}}, in: 2017 {{IEEE Conference}}
  on {{Computer Vision}} and {{Pattern Recognition}} ({{CVPR}}), {IEEE}, 2017,
  pp. 936--944.
\newblock \href {https://doi.org/10.1109/CVPR.2017.106}
  {\path{doi:10.1109/CVPR.2017.106}}.

\bibitem{Lin2014}
T.-Y. Lin, M.~Maire, S.~Belongie, J.~Hays, P.~Perona, D.~Ramanan,
  P.~Doll{\'a}r, C.~L. Zitnick, Microsoft {{COCO}}: {{Common Objects}} in
  {{Context}}, in: European {{Conference}} on {{Computer Vision}}, Vol. 8693 of
  {{LNCS}}, 2014, pp. 740--755.
\newblock \href {https://doi.org/10.1007/978-3-319-10602-1\_48}
  {\path{doi:10.1007/978-3-319-10602-1\_48}}.

\bibitem{Brockman2016}
G.~Brockman, V.~Cheung, L.~Pettersson, J.~Schneider, J.~Schulman, J.~Tang,
  W.~Zaremba, {{OpenAI Gym}}, arXiv:1606.01540 [cs] (Jun. 2016).

\bibitem{Dud2011}
M.~Dud, J.~Langford, Doubly {{Robust Policy Evaluation}} and {{Learning}}, in:
  Proceedings of the 28th {{International}} {{Conference}} on {{Machine
  Learning}}, 2011, pp. 1097--1104.

\bibitem{SharifRazavian2014}
A.~Sharif~Razavian, H.~Azizpour, J.~Sullivan, S.~Carlsson, {{CNN Features
  Off}}-the-{{Shelf}}: {{An Astounding Baseline}} for {{Recognition}}, in:
  Proceedings of the {{IEEE Conference}} on {{Computer Vision}} and {{Pattern
  Recognition Workshops}}, 2014, pp. 806--813.

\bibitem{Goodfellow2009}
I.~Goodfellow, H.~Lee, Q.~V. Le, A.~Saxe, A.~Y. Ng, Measuring {{Invariances}}
  in {{Deep Networks}}, in: Advances in {{Neural Information Processing
  Systems}}, Vol.~22, 2009, pp. 646--654.

\bibitem{Carlini2017}
N.~Carlini, D.~Wagner, Towards {{Evaluating}} the {{Robustness}} of {{Neural
  Networks}}, in: {{IEEE Symposium}} on {{Security}} and {{Privacy}}, 2017, pp.
  39--57.
\newblock \href {https://doi.org/10.1109/SP.2017.49}
  {\path{doi:10.1109/SP.2017.49}}.

\bibitem{Biggio2018}
B.~Biggio, F.~Roli, Wild patterns: {{Ten}} years after the rise of adversarial
  machine learning, Pattern Recognition 84 (2018) 317--331.
\newblock \href {https://doi.org/10.1016/j.patcog.2018.07.023}
  {\path{doi:10.1016/j.patcog.2018.07.023}}.

\end{thebibliography}

\end{document}